\documentclass[reprint,aps,showkeys,showpacs]{revtex4-1}
\usepackage{hyperref}
\usepackage{latexsym}
\usepackage{amsfonts}
\usepackage{times}
\usepackage{color}
\usepackage{url}
\usepackage{graphicx}
\usepackage{placeins}
\usepackage[noend]{algpseudocode}
\usepackage{newfloat}
\usepackage{natbib}

\DeclareFloatingEnvironment[
    fileext=loa,
    listname=List of Algorithms,
    name=ALGORITHM,
    placement=tbhp,
]{algorithm}

\newcommand{\sep}{; }
\newcommand{\remark}[1]{\ifodd\value{page} \normalmarginpar
 \else \reversemarginpar \fi \marginpar{{\footnotesize #1}} }

\newcommand{\keyw}[1]{\emph{#1}}

\newcommand{\RR}{\mathbb{R}}
\newcommand{\NN}{\mathbb{N}}
\newcommand{\ZZ}{\mathbb{Z}}

\newcommand{\network}[1]{\mathcal{#1}}
\newcommand{\vertices}[1]{\mathcal{#1}}
\newcommand{\edges}[1]{\mathcal{#1}}

\newcommand{\Net}{\network{N}}
\newcommand{\functions}[1]{\mathcal{#1}}
\newcommand{\Time}{\mathcal{T}}
\newcommand{\func}[1]{\textit{#1}}
\newcommand{\funi}[1]{{\scriptsize\textit{#1}}}
\newcommand{\cmdkey}{\raisebox{-.025em}{\includegraphics[height=.7em]{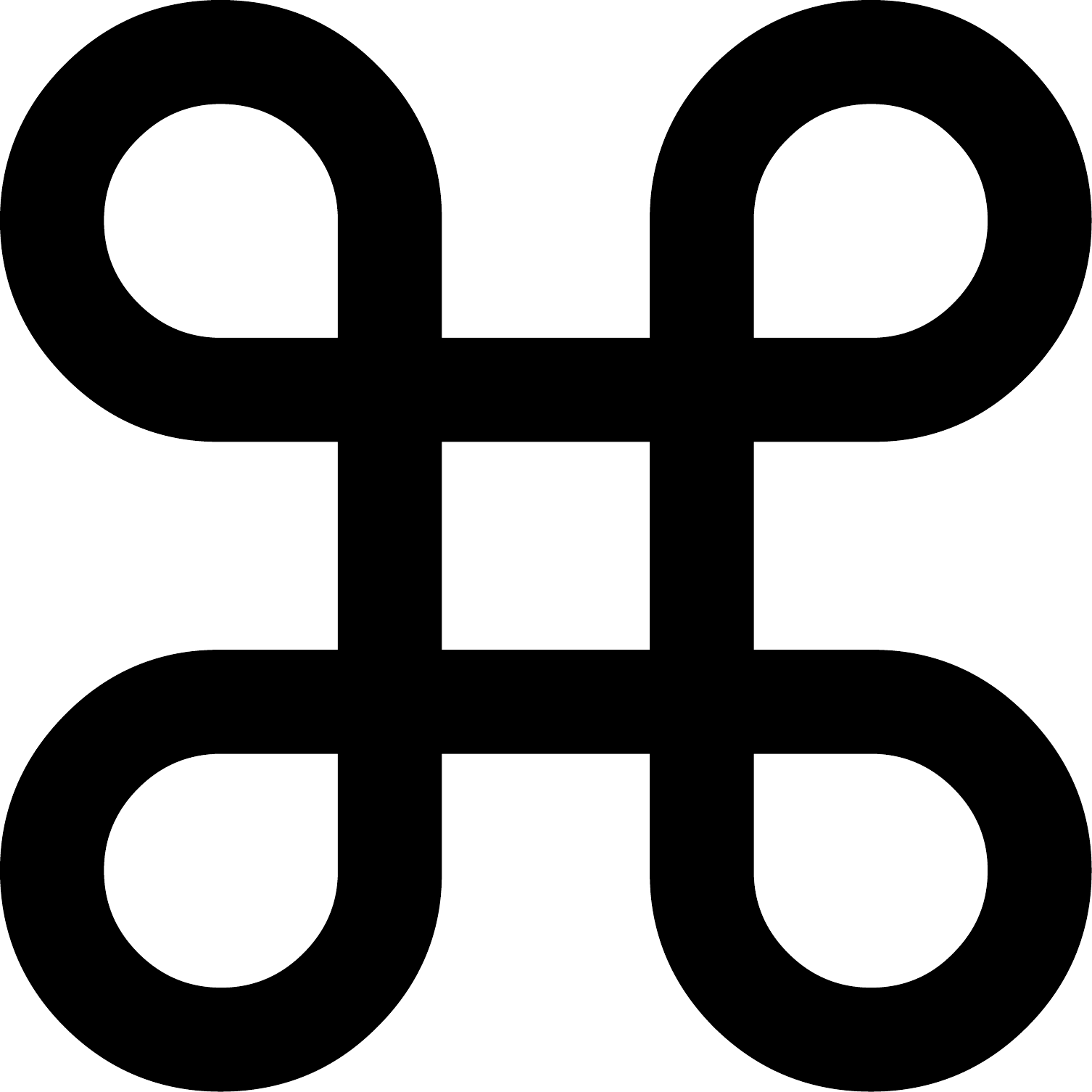}}}
\newcommand{\Mw}{\mathop{\raisebox{-1.5pt}{\mbox{$\Box$\kern-.55em\raisebox{2.5pt}{{\tiny $r$}}\kern2.9pt}}}}
\newcommand{\Mv}{\mathop{\raisebox{-1.5pt}{\mbox{$\Box$\kern-.55em\raisebox{2.5pt}{{\tiny $h$}}\kern2.9pt}}}}

\graphicspath{{./pics/}}
\date{\today}

\begin{document}

\title{An algebraic approach to temporal network analysis based on temporal quantities}
\author{Vladimir Batagelj}
\email{vladimir.batagelj@fmf.uni-lj.si}
\author{Selena Praprotnik}
\email{selena.praprotnik@gmail.com}
\affiliation{University of Ljubljana, Faculty of Mathematics and Physics, Jadranska ulica 19, 1000 Ljubljana, Slovenia}

\begin{abstract}
In a temporal network, the presence and activity
of nodes and links can change through time.
To describe temporal networks  we introduce the notion of temporal
quantities. We define the addition and multiplication of temporal quantities in
a way that can be used for the definition of addition and multiplication of
temporal networks. The corresponding algebraic structures are semirings. 
The usual approach to (data) analysis of temporal networks is to transform it into
a sequence of time slices -- static networks corresponding to selected time intervals
and analyze each of them using standard methods to produce a sequence of results.
The approach proposed in this paper enables us to compute these results directly.
We developed fast algorithms for the
proposed operations. They are available as an open source Python library TQ 
(Temporal Quantities) and a program Ianus. 
The proposed approach enables us to treat as temporal quantities also other network characteristics such as degrees, connectivity components, centrality measures, Pathfinder skeleton, etc.
To illustrate the developed tools we present some results from the analysis 
of Franzosi's violence network and Corman's Reuters terror news network.
\end{abstract}


\maketitle


\section{Introduction}

In a \keyw{temporal network}, the presence and activity
of nodes and links can change through time.
In the last two decades the interest for the analysis of temporal networks increased 
partially motivated by travel-support services and the analysis of sequences of interaction
events (e-mails, news, phone calls, collaboration, etc.). The approaches and results were recently
surveyed by Holme and Saram\"{a}ki in their paper \citep{TNsur} and the book \citep{TNbook}. 

Most of temporal social networks data contain the information about activity time
intervals of their links, sometimes augmented by the activity intensity.
The usual approach to the (data) analysis of temporal networks is to transform it into
a sequence of time slices -- static networks corresponding to selected time intervals
-- see for example \cite{DNvis,trends,elDyn}. Afterward each time slice is analyzed using the standard methods for analysis of
static networks. Finally the results are collected into a temporal sequence of results.
In this paper we propose an alternative approach, based on the notion of temporal
quantity, that bypasses explicit construction of time slices. The developed 
algorithms are transforming temporal networks directly into results in the form of
temporal quantities, vectors, temporal vectors or partitions, and temporal networks.

In the paper, we first present the basic notions about temporal networks.
In Section~\ref{secTQ} we introduce the temporal quantities and propose an
algebraic approach, based on semirings, to the analysis of temporal networks.
In the following sections we show that most of the traditional network analysis
concepts and algorithms such as degrees, clustering coefficient, closeness,
betweenness, weak and strong connectivity, PathFinder skeleton, etc. can be
straightforwardly extended to their temporal versions.

\section{Description of temporal networks \label{desc}}

For the description of temporal networks we propose an elaborated version of the
approach used in Pajek \citep{ESNA}. In our approach we also consider values of links (in most cases
measuring the intensity/frequency of the activity).
Pajek supports two types of descriptions of temporal
networks based on \keyw{presence} and on \keyw{events} (Pajek 0.47, July 1999).
Here, we will describe only the approach to capturing the presence of nodes and links.

A \keyw{temporal network}
$\network{N}_T =(\vertices{V},\edges{L}, \Time,\functions{P},\functions{W})$
is obtained by attaching the \keyw{time}, $\Time$, to an ordinary network, where
$\Time$ is a set of  \keyw{time points}, $t \in \Time$.
$\vertices{V}$ is the set of nodes, $\edges{L}$ is the set of links, $\functions{P}$
is the set of node properties, and $\functions{W}$ is the set of link properties or  weights
\citep{ency}. The time $\Time$ is usually either a subset of integers,  
$\Time \subseteq \ZZ$, or a subset of reals,  $\Time \subseteq \RR$.
In Pajek $\Time \subseteq \NN$. In a general setting it could be any linearly ordered
set.

In a temporal network, nodes $v \in \vertices{V}$ and links $l \in \edges{L}$
are not necessarily present or active at all time points.
Let $T(v)$, $T \in \functions{P}$, be the activity set of time points for the
node $v$; and  $T(l)$, $T \in \functions{W}$, the activity set of time points
for the link $l$.
The following \keyw{consistency} condition is imposed:
If a link $l(u,v)$ is active at the time point $t$ then
its end-nodes $u$ and $v$ should be active at the time $t$.
Formally we express this by 
\[ T(l(u,v)) \subseteq T(u) \cap T(v) . \]
The activity set $T(e)$ of a node/link $e$ is usually described as a sequence
of activity time intervals $([s_i,f_i))_{i=1}^k$, where $s_i$ is the
\keyw{start}ing time and $f_i$ is the \keyw{finish}ing time.
 
We denote a network consisting of links and
nodes active in the time $t \in \Time$ by $\network{N}(t)$
and call it the (network) \keyw{time slice} or \keyw{footprint} of $t$.
Let $\Time' \subset \Time$ (for example, a time interval). The notion
of a time slice is extended to $\Time'$ by
\[ \network{N}(\Time') = \bigcup_{t\in \Time'} \network{N}(t) . \]
 
\subsection{Examples}

Let us look at some examples of temporal networks.

\textbf{Citation networks} can be obtained from bibliographic data bases such as Web of Science (Knowledge) and Scopus. In a citation network $\network{N} =(\vertices{V},\edges{L}, \Time,\functions{P},\functions{W})$,
its set of nodes $\vertices{V}$  consists of selected works (papers, books, reports, patents, etc.). There exists
an arc $a(u,v) \in \edges{L}$ iff  the  work $u$ cites the work $v$. The time set $\Time$ is usually an interval of
years $[\func{year}_\funi{first}, \func{year}_\funi{last}]$ in which the works were published. The activity set of the work $v$, $T(v)$, is the interval $[\func{year}_\funi{pub}(v), \func{year}_\funi{last}]$; and the activity set of the arc $a(u,v)$, $T(a)$, can be set to the interval 
$[\func{year}_\funi{pub}(u), \func{year}_\funi{pub}(u)]$ (instances approach) or to the interval
$[\func{year}_\funi{pub}(u), \func{year}_\funi{last}]$ (cumulative approach). An example of a property $p \in \functions{P}$ is the number of pages or the number of authors.
Other properties, such as work's authors and keywords, are usually represented as two-mode networks.

\textbf{Project collaboration networks} are usually based on some project data base such as Cordis. 
The set of nodes $\vertices{V}$ consists of participating institutions. There is an edge $e(u\colon v) \in \edges{L}$
iff institutions $u$  and  $v$ work on a joint project. The time set $\Time$ is an
interval of dates/days $[\func{day}_\funi{first}, \func{day}_\funi{last}]$ in which the collaboration data
were collected.
$T(v) = \Time$ and
$T(e) = \{  [s,f] : $ there exists a project $P$ such that $u$ and $v$ are
 partners on $P$;  $s$ is the start and $f$ is the finish date of $P \}$.

\textbf{KEDS/WEIS networks} are networks registering political events in critical
regions in the world (Middle East, Balkans, and West Africa) on the basis of
daily news. Originally they were collected by KEDS (Kansas Event Data System).
Currently they are hosted by Parus Analytical Systems.
The set of nodes  $\vertices{V}$ contains the involved actors (states, political
groups, international organizations, etc.). The links are directed and are
describing the events: 
\[ ( date, actor_1, actor_2, action ) \]
on a given $date$ the $actor_1$ made the $action$ on the $actor_2$. Different actions are 
determining different relations -- we get a multirelational network
with a set of links partitioned by actions
$\edges{L} = \{ \edges{L}_a : a \in \func{Actions} \}$.
The time set is determined by the observed period $\Time = [day_\funi{first},day_\funi{last}]$.
Since most of the actors are existing during all the observed period their
node activity time sets are $T(v) = \Time$. Another option is to consider  
as their node activity time sets the period of their engagement in the region.
The activity time set $T(l)$ of an arc $l(u,v) \in \edges{L}_a$ contains all
dates -- intervals $[day,day+1)$ -- in which the actor $u$ made an action $a$ on
the actor $v$. Another possibility is to base the description on a single relation
network and store the information about the action $a$ as a structured value in
a triple $(day,day+1,value)$
\[ value = [(action_1,count_1), \ldots ,(action_k,count_k) ] \]
and introduce an appropriate semiring over such values (see Section~\ref{secTQ}).

There are many other examples of temporal networks such as: genealogies,
contact networks, networks of phone calls, etc.

\section{Temporal quantities}\label{secTQ}

Besides the presence/absence of nodes and links also their properties can change
through time. To describe the changes we introduce the notion of a \keyw{temporal quantity} $a$ with the activity set $T_a \subseteq \Time$
\[  a = \left\{\begin{array}{ll} 
                a'(t) & t \in T_a \\
                \cmdkey & t \in \Time \setminus T_a
             \end{array}\right. \]
where $a'(t)$ is the value of $a$ at an instant $t$, and  \cmdkey{} denotes the
value \keyw{undefined}.

We assume that the values of temporal quantities belong to a set $A$ which is
a semiring $(A,\oplus,\odot,0,1)$ for binary operations $\oplus : A\times A \to A$ and
$\odot : A\times A \to A$ \citep{GoMi,semi}. This means that $(A,\oplus,0)$ is an Abelian monoid -- the
addition $\oplus$ is associative and commutative, and has 0 as its neutral element;
and  $(A,\odot,1)$ is a monoid -- the multiplication $\odot$ is associative 
and has 1 as its neutral element. Also, multiplication distributes from both
sides over the addition.
Note that $0$ and $1$ denote the two elements of $A$ that satisfy the required properties. 
In expressions the precedence of the multiplication $\odot$ over the addition
$\oplus$ is assumed. We can extend both operations to the set
$A_{\scriptsize\cmdkey} = A \cup \{\cmdkey\}$ by requiring that for all $a \in A_{\scriptsize\cmdkey}$
it holds
\[ a \oplus \cmdkey = \cmdkey \oplus a = a \quad \mbox{and} \quad
   a \odot \cmdkey = \cmdkey \odot a = \cmdkey . \]
The structure $(A_{\scriptsize\cmdkey},\oplus,\odot,\cmdkey,1)$ is also a semiring.

\begin{figure}[!]
 \begin{center}
   \includegraphics[width=75mm]{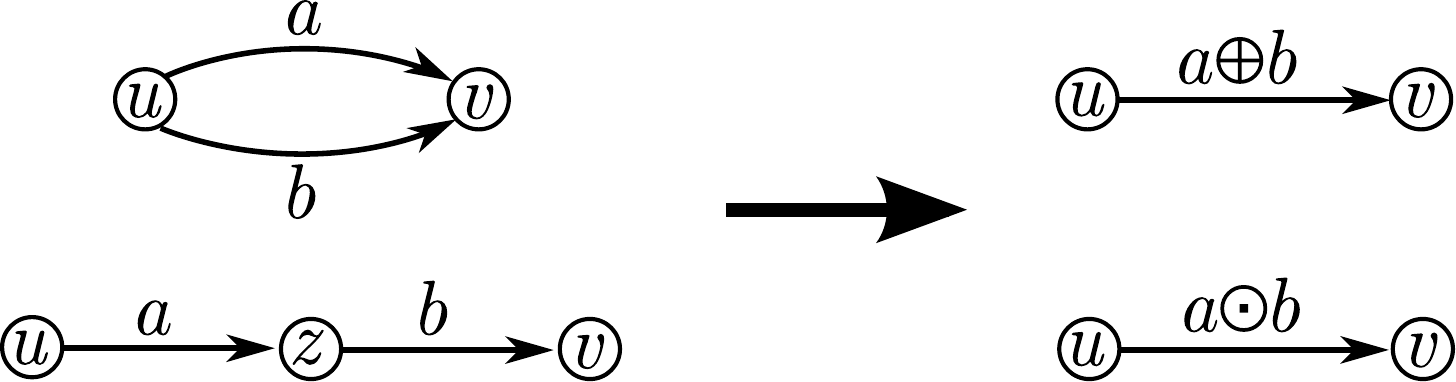}
   \caption{Semiring addition and multiplication in networks.\label{semiG}}
 \end{center}
\end{figure}

The ``default'' semiring is the \keyw{combinatorial} semiring $(\RR_0^+,+,\cdot,0,1)$
where $+$ and $\cdot$ are the usual addition and multiplication of real
numbers. In some applications other semirings are useful.

In applications of semirings in the analysis of graphs and networks the addition $\oplus$
describes the composition of values on parallel walks and the multiplication $\odot$
describes the composition of values on sequential walks -- see Figure~\ref{semiG}.
For the combinatorial semiring these two schemes correspond to basic principles
of combinatorics: the \keyw{Rule of Sum} and the \keyw{Rule of Product} \citep{comb}.

The semiring $(\overline{\RR_0^+}, \min, +, \infty, 0)$, $\overline{\RR_0^+} 
= \RR_0^+ \cup \{\infty\}$, is suitable to deal
with the shortest paths problem in networks; and the semiring $(\{0,1\}, \lor, 
\land, 0, 1)$ for reachability problems. The standard references on semirings
are \cite{GaN} and \cite{GoMi}.

\subsection{Semiring of temporal quantities}

Let $A_{\scriptsize\cmdkey}(\Time)$ denote the set of all temporal quantities
over $A_{\scriptsize\cmdkey}$ in the time $\Time$. To extend the operations to
networks and their matrices we first define the \keyw{sum} (parallel links)
$ a \oplus b $ as
\[  (a \oplus b)(t) =  a(t) \oplus b(t) \]
and $T_s = T_a \cup T_b$; and the \keyw{product} (sequential links) 
$ a \odot b $ as             
\[ (a \odot b)(t)  = a(t) \odot b(t) \]
and $T_p = T_a \cap T_b$. 

In these definitions and also in the following text, to avoid the `pollution' with many
different symbols, we use the symbols $\oplus$ and $\odot$ to denote the semiring operations.
The appropriate semiring can be determined from the context. For example, in the definition
of addition of temporal quantities the symbol $\oplus$ on the left hand side of the equation operates
on temporal quantities and the symbol $\oplus$ on the right hand side denotes the
addition in the basic semiring $A_{\scriptsize\cmdkey}$.

Let us define the temporal quantities $\mathbf{0}$ and $\mathbf{1}$ with
requirements $\mathbf{0}(t) = \cmdkey$ and $\mathbf{1}(t) = 1$ for all
$t \in \Time$. It is a routine task to verify that the structure 
$(A_{\scriptsize\cmdkey}(\Time),\oplus,\odot,\mathbf{0},\mathbf{1})$ is also a 
semiring, and therefore so is the set of square matrices of order $n$ over it for the
addition $\mathbf{A} \oplus \mathbf{B} = \mathbf{S}$
\[ s_{ij} = a_{ij} \oplus b_{ij} \]
and multiplication $\mathbf{A} \odot \mathbf{B} = \mathbf{P}$            
\[ p_{ij} = \bigoplus_{k=1}^n a_{ik} \odot b_{kj} . \]
Again, the symbols $\oplus$ and $\odot$  on the left hand side operate on temporal 
matrices and on the right hand side in the semiring of temporal quantities.

The matrix multiplication is closely related to traveling on networks. Consider an
entry $p_{ij}$ in an instant $t$
\[ p_{ij}(t) = \bigoplus_{k=1}^n a_{ik}(t) \odot b_{kj}(t) . \]
For a value $p_{ij}(t)$ to be defined (different from $\cmdkey$) there should exist
in the instant $t$ at least
one node $k$ such that both the link $(i,k)$ and the link $(k,j)$ exist -- the transition 
from the node $i$ to the node $j$ through a node $k$ is possible. Its contribution is
$a_{ik}(t)\odot b_{kj}(t)$. This means that the matrix multiplication is taking into
account only the links inside the time slice $\network{N}(t)$.

\subsection{Operationalization}

In the following we shall limit our discussion to temporal quantities that can be
described in the form of time-interval/value sequences
\[ a = ( (I_i, v_i) )_{i=1}^k \]
where $I_i$ is a time-interval and $v_i$ is a value of $a$ on this interval. In general, the
intervals can be of different types: 1 -- $[ s_i, f_i]$;
2 -- $[ s_i, f_i)$; 3 -- $( s_i, f_i]$; 4 -- $( s_i, f_i)$. Also the value $v_i$ 
can be structured. For example $v_i = (w_i, c_i, \tau_i)$ -- weight, capacity and
transition time, or $v_i = (d_i,n_i)$ -- the length of geodesics and the number of geodesics,
etc. We require 
$s_i \leq f_i$, for $i=1,\ldots,k$ and $s_{i-1} < s_i$, for $i=2,\ldots,k$.

\begin{algorithm}
\caption{Addition of temporal quantities. }
\label{sumTQ}
\begin{algorithmic}[1]
\Function{\func{sum}}{$a,b$}
   \If{$\func{length}(a) = 0$} \Return  $b$ \EndIf
   \If{$\func{length}(b) = 0$} \Return $a$ \EndIf
   \State $c \gets [\ ]$
   \State $(s_a,f_a,v_a) \gets \func{get}(a)$; $(s_b,f_b,v_b) \gets \func{get}(b)$
   \While{$(s_a<\infty) \lor (s_b<\infty)$}
      \If{$s_a < s_b$}
         \State $s_c \gets s_a$; $v_c \gets v_a$
         \If {$s_b < f_a$} $f_c \gets s_b$; $s_a \gets s_b$
         \Else $f_c \gets f_a$; $(s_a,f_a,v_a) \gets \func{get}(a)$ \EndIf
      \ElsIf{$s_a = s_b$}
         \State $s_c \gets s_a$; $f_c \gets \min(f_a,f_b)$
         \State $v_c \gets \func{sAdd}(v_a,v_b)$
         \State $s_a \gets s_b \gets f_c$; $f_d \gets f_a$
         \If{$f_d \leq f_b$} $(s_a,f_a,v_a) \gets \func{get}(a)$ \EndIf
         \If{$f_b \leq f_d$} $(s_b,f_b,v_b) \gets \func{get}(b)$ \EndIf
      \Else
         \State $s_c \gets s_b$; $v_c \gets v_b$
         \If{$s_a < f_b$} $f_c \gets s_a$; $s_b \gets s_a$
         \Else $f_c \gets f_b$; $(s_b,f_b,v_b) \gets \func{get}(b)$ \EndIf
      \EndIf
      \State $c.\func{append}((s_c,f_c,v_c))$
   \EndWhile   
   \State \Return $\func{standard}(c)$
\EndFunction
\end{algorithmic}
\end{algorithm}

To simplify the exposition we will assume in the following that all the intervals in
our descriptions of temporal quantities are of type 2 -- $[ s_i, f_i)$  and $f_{i-1} \leq s_i$, for $i=2,\ldots,k$. Therefore
we can describe the temporal quantities with sequences of triples
\[ a = ( (s_i, f_i, v_i) )_{i=1}^k . \]
In the examples we will also assume that ${\cal T} = [t_{min}, t_{max}] \subset \NN$.

To provide a computational support for the proposed approach we are developing
in Python a library TQ (Temporal Quantities). In the examples  we will use the
Python notation for temporal quantities.
 
The following are two temporal quantities $a$ and $b$ represented in Python as a
list of triples

{\renewcommand{\baselinestretch}{0.8}\small
\begin{verbatim}
a = [(1, 5, 2), (6, 8, 1), (11, 12, 3),
    (14, 16, 2), (17, 18, 5), (19, 20, 1)]
b = [(2, 3, 4), (4, 7, 3), (9, 10, 2), 
    (13, 15, 5), (16, 21, 1)]
\end{verbatim}\normalsize
}
The temporal quantity $a$ has on the interval $[1,5)$ (i.e. in instances 1, 2, 3 and 4) value 2; on the interval $[6,8)$
value 1; on the interval $[11,12)$ value 3, etc. Outside the specified intervals
its value is undefined, $\cmdkey$.

\begin{figure}
 \begin{center}
    \begin{tabular}{l}
   $a$ :\\
   \includegraphics[width=75mm,viewport=140 80 580 220,clip=]{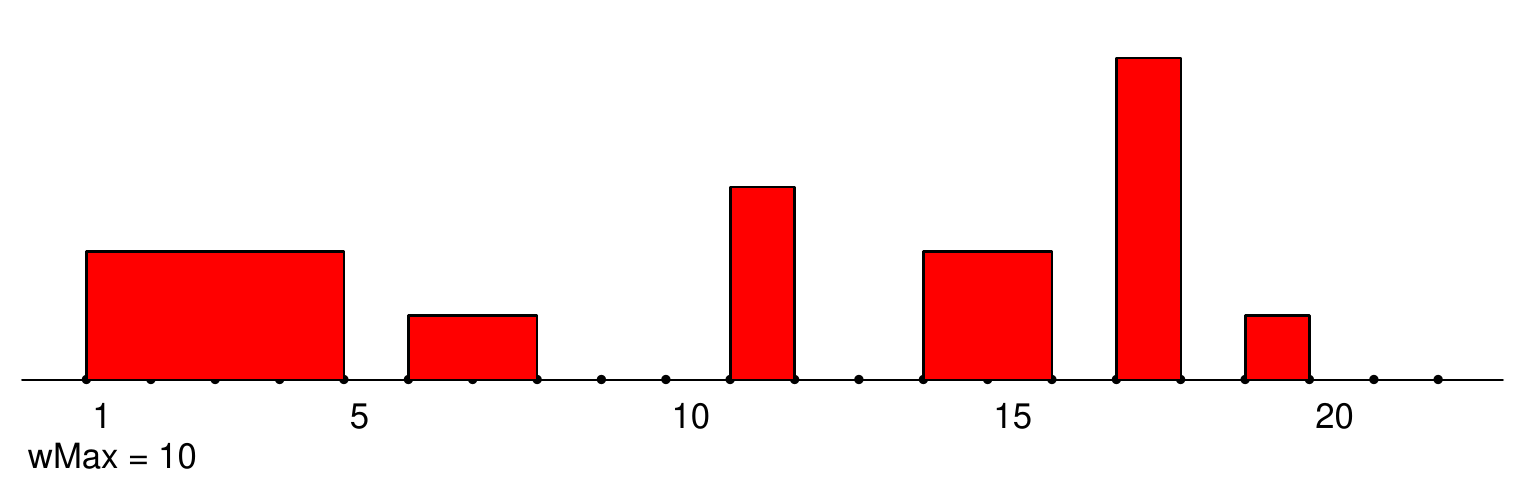}\\
  $b$ :\\ 
  \includegraphics[width=75mm,viewport=140 80 580 220,clip=]{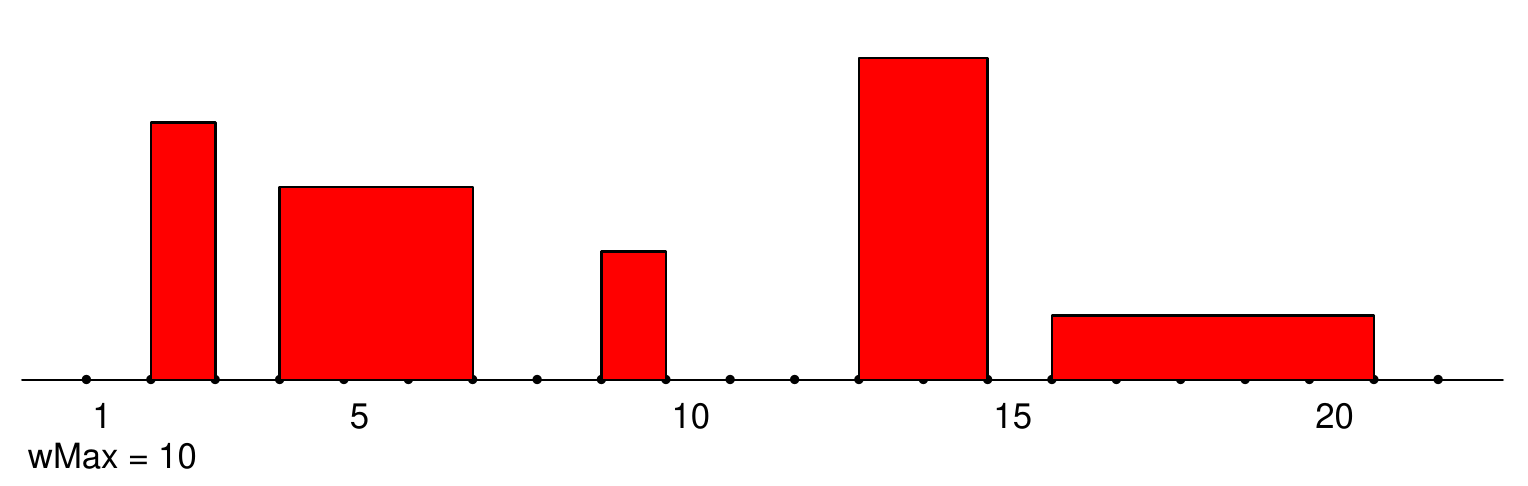}\\
  $a\oplus b$ :\\
   \includegraphics[width=75mm,viewport=140 80 580 260,clip=]{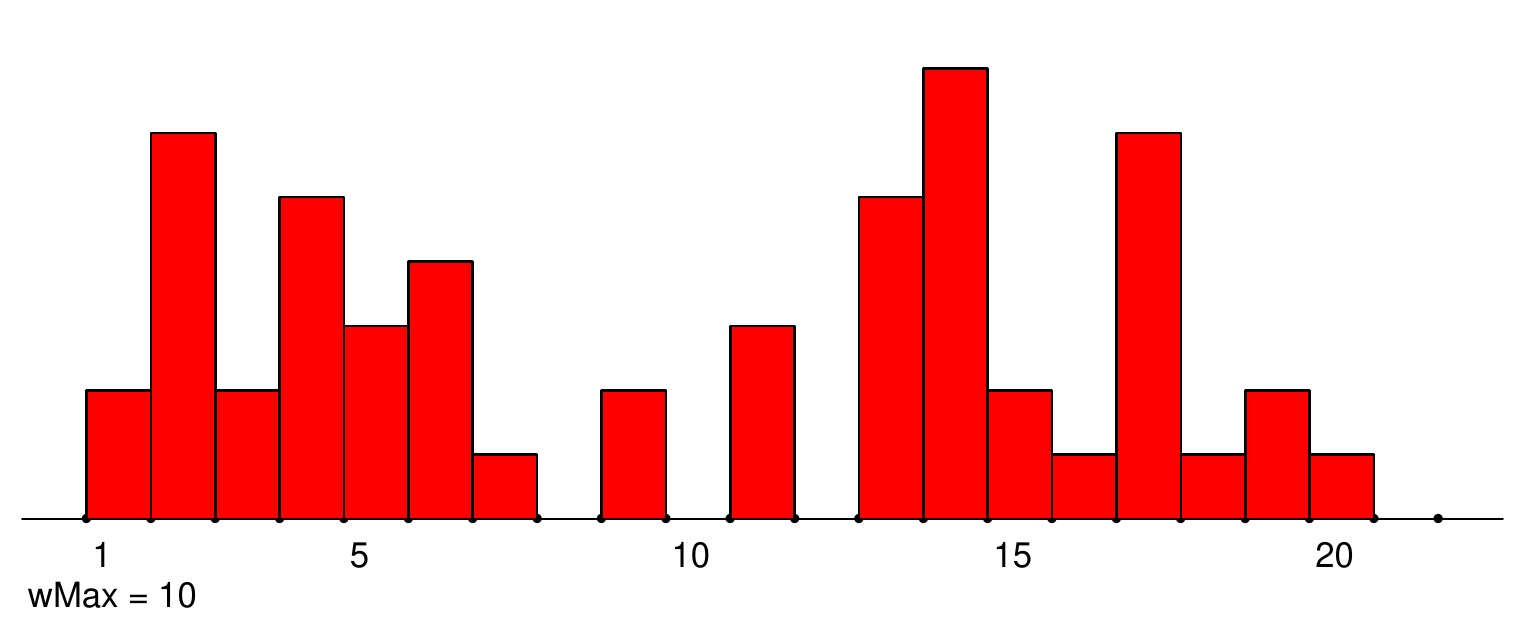}\\
   $a\odot b$ :\\
   \includegraphics[width=75mm,viewport=140 80 580 305,clip=]{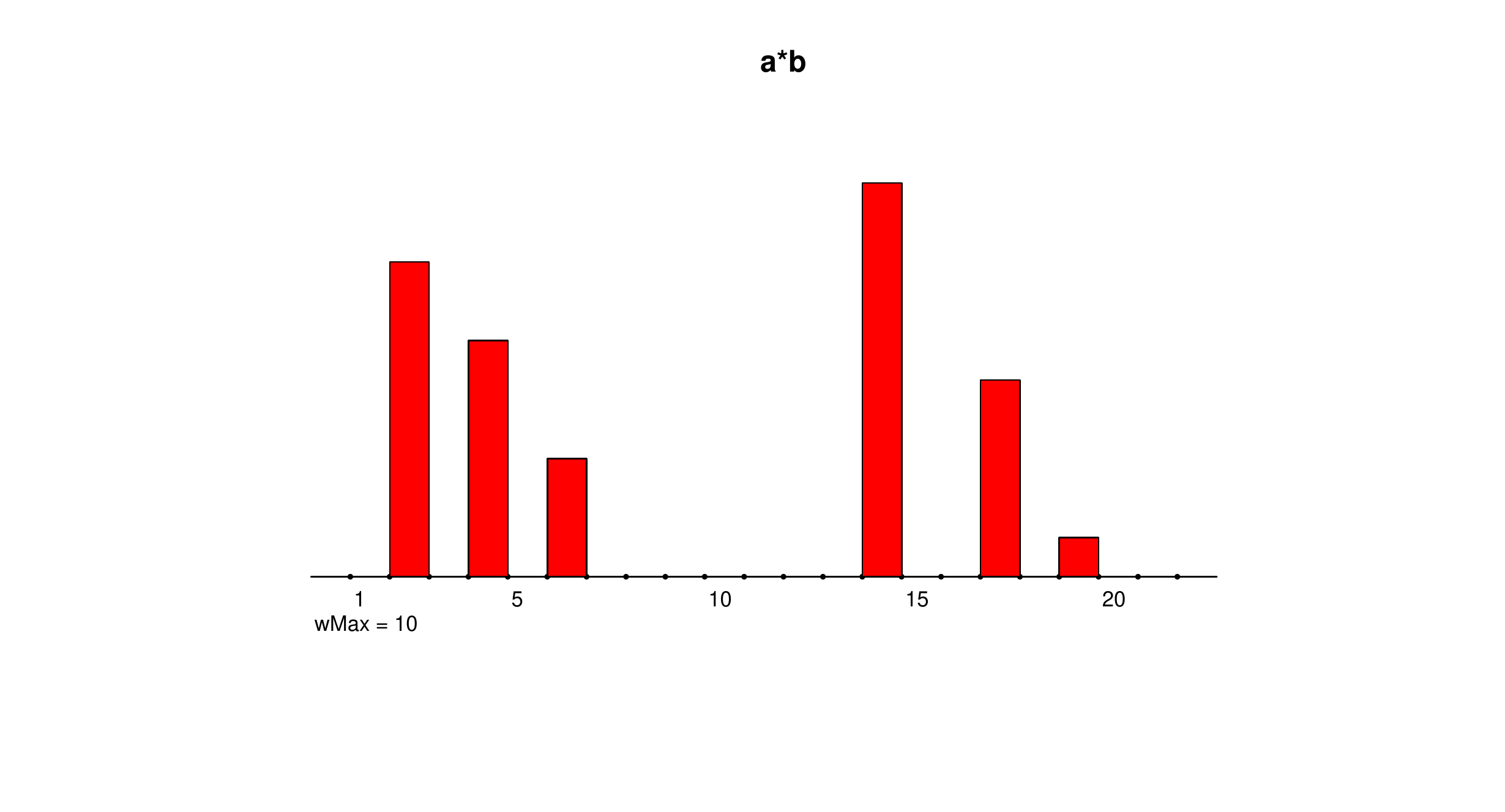}
    \end{tabular}
  \caption{Addition and multiplication of temporal quantities.\label{ops}}
 \end{center}
\end{figure}

The temporal quantities can also be visualized as it is shown for $a$ and $b$ at
the top half of Figure~\ref{ops}.

For the simplified version of temporal quantities we wrote procedures $\func{sum}$
(Algorithm~\ref{sumTQ}) for the addition and $\func{prod}$ (Algorithm~\ref{proTQ}) for the
multiplication of temporal quantities over the selected semiring. Because, by assumption,
the triples in a description of a temporal quantity are ordered by their starting
times, we can base both procedures on the ordered lists merging scheme. The basic semiring
operations of addition and multiplication are provided by functions $\func{sAdd}$ and $\func{sMul}$.

The function $\func{length}(a)$ returns the length (number of items) of the list $a$.
The function $\func{get}(a)$ returns the current item of the list $a$ and moves to the next
item; if the list is exausted it returns a `sentinel' triple $(\infty,\infty,0)$.
The statement $(s,f,v) \gets e$ describes the unpacking of the item $e$ into its parts.
The statement $c.\func{append}(e)$ appends the item $e$ to the tail of the list $c$.
The function $\func{standard}(a)$ joins, in the list $a$, adjacent time intervals with the same
value into a single interval.

\begin{algorithm}
\caption{Multiplication of temporal quantities. }
\label{proTQ}
\begin{algorithmic}[1]
\Function{\func{prod}}{$a,b$}
   \If{$\func{length}(a)\cdot \func{length}(b) = 0$} \Return $[\ ]$ \EndIf
   \State $c \gets [\ ]$; $(s_a,f_a,v_a) \gets \func{get}(a)$; $(s_b,f_b,v_b) \gets \func{get}(b)$
   \While{$(s_a<\infty) \lor (s_b<\infty)$}
      \If{$f_a \leq s_b$} $(s_a,f_a,v_a) \gets \func{get}(a)$  
      \ElsIf{$f_b \leq s_a$} $(s_b,f_b,v_b) \gets \func{get}(b)$ 
      \Else
         \State $s_c \gets \max(s_a,s_b)$; $f_c \gets \min(f_a,f_b)$;
         \State $v_c \gets \func{sMul}(v_a,v_b)$
         \State $c.\func{append}((s_c,f_c,v_c))$
         \If{$f_c = f_a$} $(s_a,f_a,v_a) \gets \func{get}(a)$ \EndIf
         \If{$f_c = f_b$} $(s_b,f_b,v_b) \gets \func{get}(b)$ \EndIf
      \EndIf
   \EndWhile   
   \State \Return $\func{standard}(c)$
\EndFunction

\end{algorithmic}
\end{algorithm}

\begin{figure}
 \begin{center}
    \begin{tabular}{l}
    $a$ :\\
   \includegraphics[width=75mm,viewport=110 112 790 160,clip=]{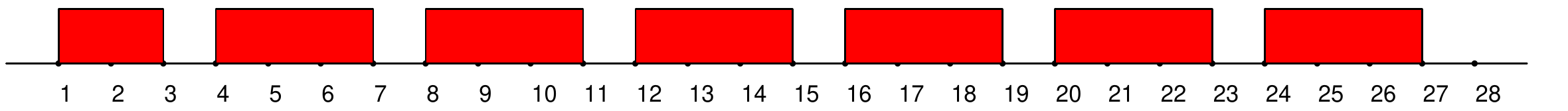}\\
   $b$ :\\
  \includegraphics[width=75mm,viewport=110 110 790 180,clip=]{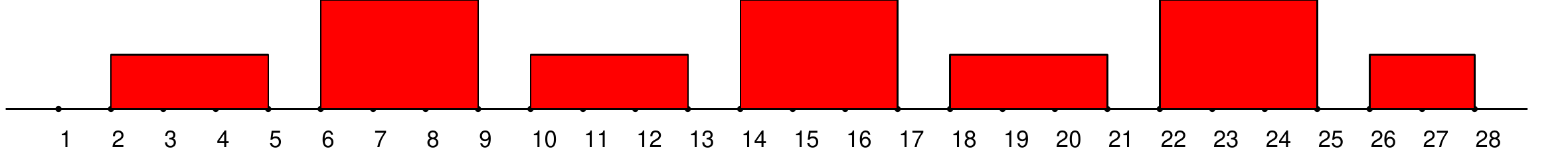}\\
  $a\oplus b$ :\\
   \includegraphics[width=75mm,viewport=110 112 790 205,clip=]{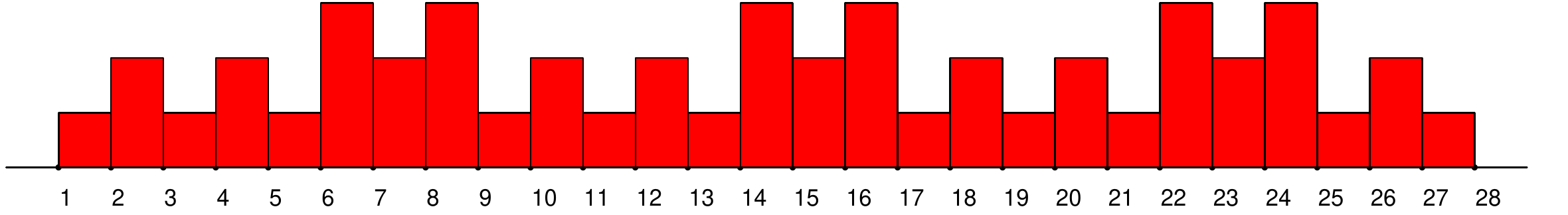}\\
   $a\odot b$ :\\
   \includegraphics[width=75mm,viewport=110 110 790 180,clip=]{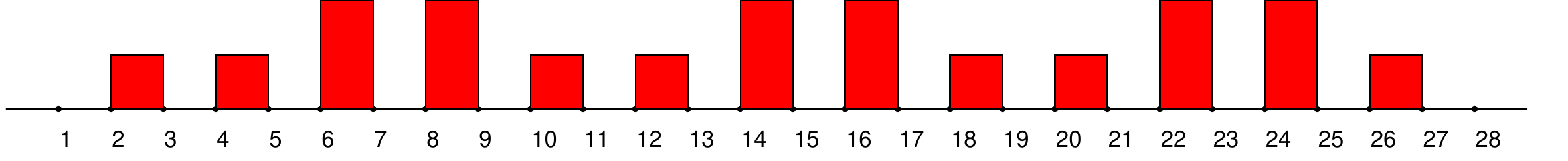}
    \end{tabular}
  \caption{Addition and multiplication of temporal quantities -- growth of size.\label{grops}}
 \end{center}
\end{figure}

The following are the sum $s$ and the product $p$ of temporal quantities $a$ and $b$.
They are visually displayed at the bottom half of Figure~\ref{ops}.

{\renewcommand{\baselinestretch}{0.8}\footnotesize
\begin{verbatim}
s = [(1, 2, 2), (2, 3, 6), (3, 4, 2),
    (4, 5, 5), (5, 6, 3), (6, 7, 4),
    (7, 8, 1), (9, 10, 2), (11, 12, 3),
    (13, 14, 5), (14, 15, 7), (15, 16, 2),
    (16, 17, 1), (17, 18, 6), (18, 19, 1),
    (19, 20, 2), (20, 21, 1)]
p = [(2, 3, 8), (4, 5, 6), (6, 7, 3),
    (14, 15, 10), (17, 18, 5), (19, 20, 1)] 
\end{verbatim}\normalsize
}

Let $l_a = \func{length}(a)$ and $l_b = \func{length}(b)$. 
Then, assuming that the semiring operations take constant
time each, the time complexity of both algorithms is $O(l_a+l_b)$.
The example in Figure~\ref{grops} shows that in extreme cases the sum can be
almost 4 times longer than each of its arguments, and the product almost twice as long as the arguments.
If ${\cal T} = [t_{min}, t_{max}] \subset \NN$ the length of a list describing a
temporal quantity can not exceed $L = t_{max}-t_{min}$.

\subsection{The aggregated value}

In some applications over the combinatorial semiring we shall use the \keyw{aggregated value} of a temporal
quantity $a = ( (s_i, f_i, v_i) )_{i=1}^k$. It is defined as
\[ \Sigma a = \sum_{i=1}^k (f_i-s_i)\cdot v_i \]
and is computed using the procedure $\func{total}(a)$. For example
$\Sigma a = 23$ and $\Sigma b = 30$. Note that $\Sigma a + \Sigma b = 
\Sigma (a+b)$.

\subsection{Temporal partitions}

The description of temporal partitions has the same form as the
description of temporal quantities $a = ( (s_i, f_i, v_i) )_{i=1}^k$.
They differ only in the interpretation of values $v_i \in \NN$. In case
of partitions $v_i = j$ means that the unit described with $a$ belongs
to a class $j$ in the time interval $[s_i,f_i)$. We shall use temporal partitions
to describe connectivity components in Section~\ref{conn}. 

We obtain a more adequate description of temporal networks by using vectors
of temporal quantities (temporal vectors and temporal partitions) for
describing properties of nodes and making also link weights into temporal
quantities. In the current version of the library TQ we use a 
representation of a network $\Net$ with its matrix $\mathbf{A} = [a_{uv}]$
\[ a_{uv} =  \left\{\begin{array}{ll} 
                w(u,v) & (u,v) \in \edges{L} \\
                \cmdkey         & \mbox{otherwise}
             \end{array}\right.
\] 
where $ w(u,v)$ is a temporal weight attached to a link $(u,v)$.

\subsection{Products of a temporal matrix and a temporal vector}

In some applications the product of a temporal matrix with a temporal vector
is useful. There are two products -- left and right.

Let $\mathbf{A}$ be a temporal matrix of size $n\times m$, $\mathbf{v}$ a
vector of size $n$, and $\mathbf{u}$ a vector of size $m$. The 
\keyw{product from left} of $\mathbf{A}$ with $\mathbf{v}$, denoted by
$\mathbf{u} = \mathbf{v} \bullet \mathbf{A}$, is defined by
\[  u_j = \bigoplus_{i=1}^n v_i \odot a_{ij}, \qquad j = 1, \ldots, m   \]
and the \keyw{product from right} of $\mathbf{A}$ with $\mathbf{u}$, denoted by
$\mathbf{v} = \mathbf{A} \bullet  \mathbf{u}$, is defined by
\[  v_i = \bigoplus_{j=1}^m  a_{ij} \odot u_j, \qquad i = 1, \ldots, n  . \]
In the TQ library both products are implemented as functions $\func{MatVecMulL}(A,v)$ and
$\func{MatVecMulR}(A,v)$.

If a vector $\mathbf{v}$ of size $n$ is considered as a column vector -- 
an $n \times 1$ matrix --  it holds 
$\mathbf{v} \bullet \mathbf{A} = (\mathbf{v}^T \odot \mathbf{A})^T$ and
$\mathbf{A} \bullet  \mathbf{u} = \mathbf{A} \odot  \mathbf{u}$. $T$ denotes
the matrix transposition operation.

\section{Node activities\label{activ}}

In this section we show how we can use the proposed operations with temporal 
quantities (the addition) for a simple analysis of temporal networks.

Assume that the values in temporal quantities $a_{uv}$  from  a temporal network
matrix $\mathbf{A}$ are positive real numbers measuring the intensity of the
activity of the node $u$ on the node $v$.
We define the \keyw{activity} of a group of nodes $\vertices{V}_1$  on a group
$\vertices{V}_2$ (using the combinatorial semiring) as 
\[ \mbox{act}(\vertices{V}_1,\vertices{V}_2) = 
 \sum_{u\in\vertices{V}_1} \sum_{v\in\vertices{V}_2} a_{uv} . \] 
To illustrate the notion of activity we applied it on Franzosi's violence 
temporal network \citep{RF}. Roberto Franzosi collected from the journal news in the
period January 1919 -- December 1922 information about the different types
of interactions between political parties and other groups of people in Italy. 
The violence network
contains only the data about violent actions and counts the number of
interactions per month.

\begin{figure*}[!]
 \begin{center}
  \large
    \begin{tabular}{rl}
  \raisebox{10mm}{police :} &
   \includegraphics[width=100mm,viewport=160 90 1330 340,clip=]{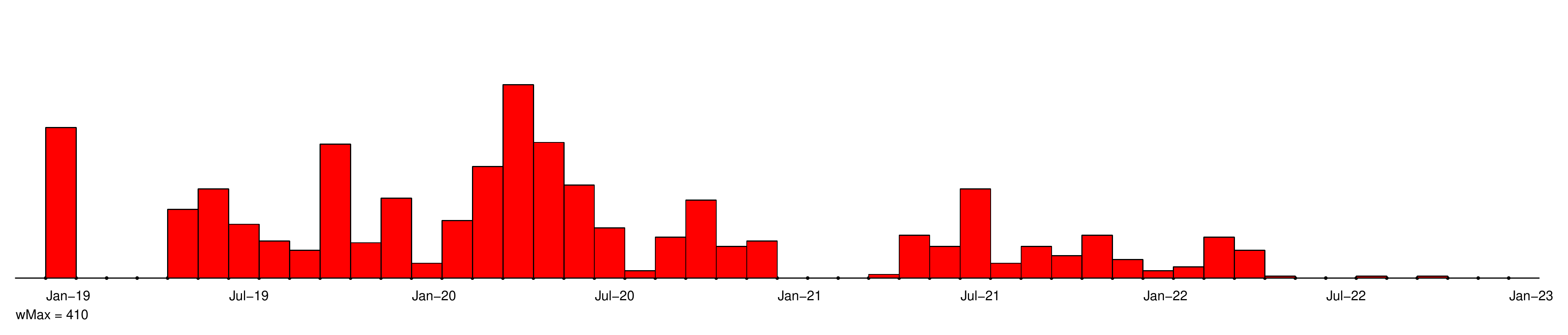}\\
  \raisebox{10mm}{fascists :} & 
  \includegraphics[width=100mm,viewport=160 90 1330 630,clip=]{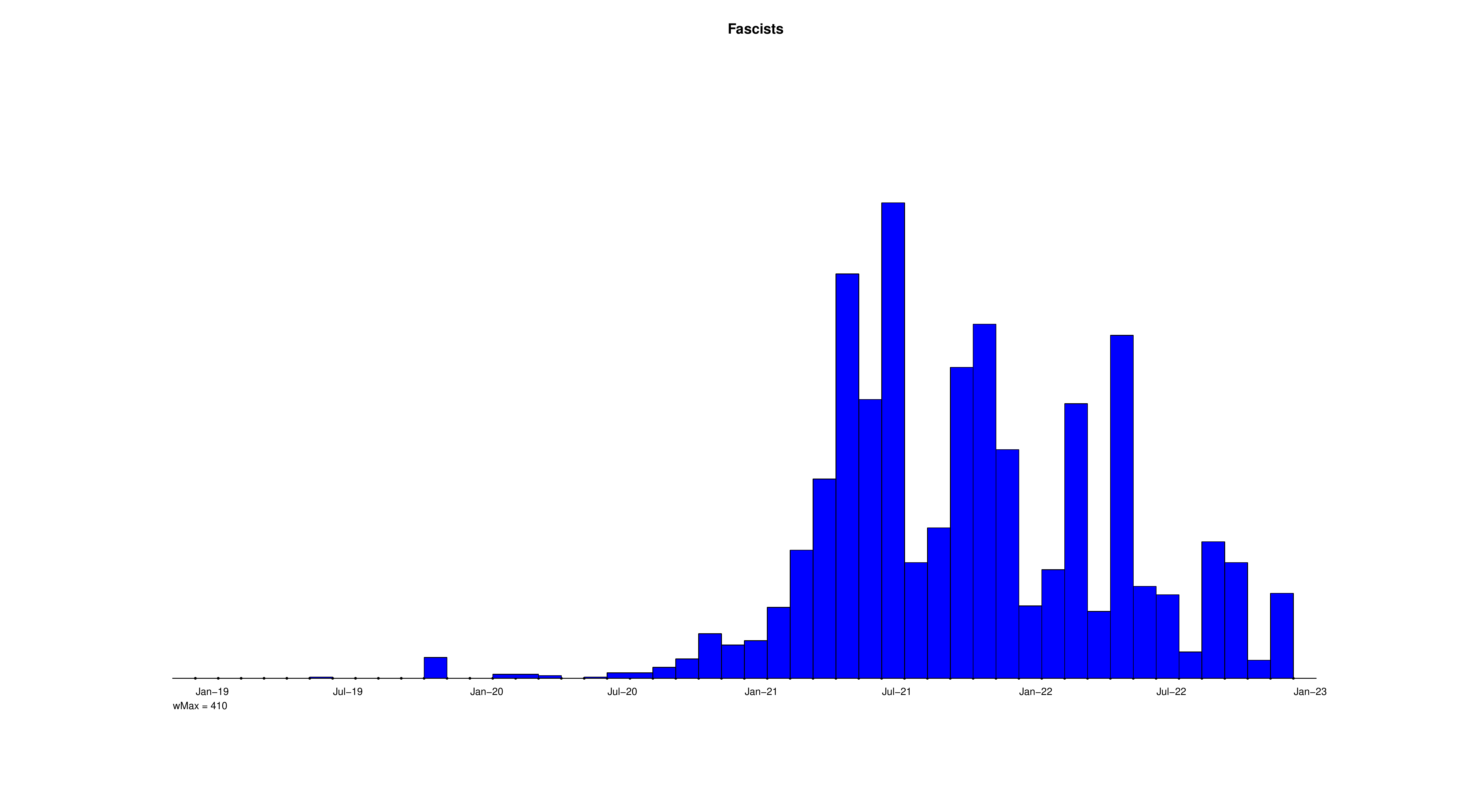}\\
  \raisebox{10mm}{all :} &
   \includegraphics[width=100mm,viewport=160 90 1330 700,clip=]{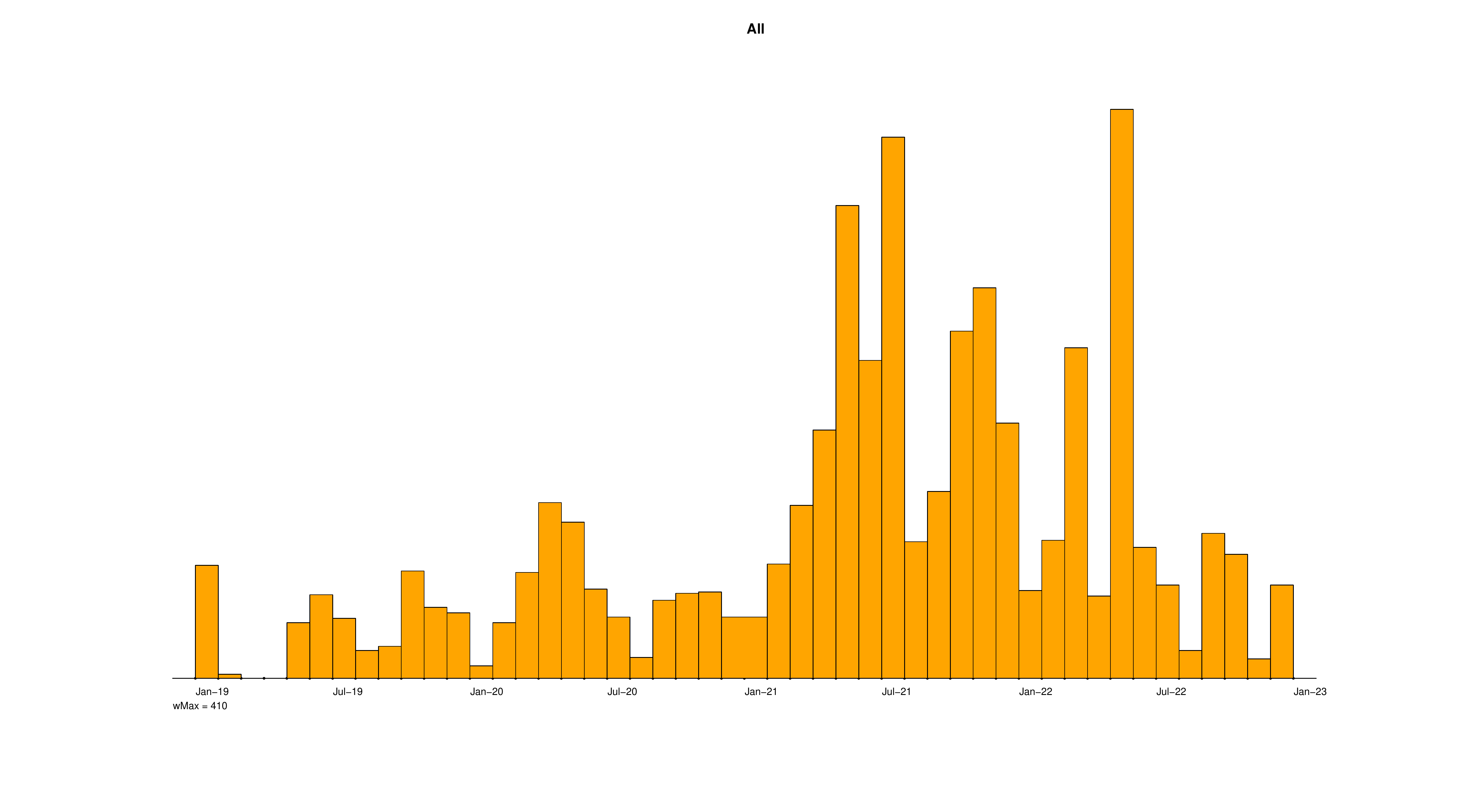}
    \end{tabular}
  \caption{Intensity of violent activities of police, fascists and all.\label{viol}}
 \end{center}
\end{figure*}

We determined the temporal quantities
 $pol = \mbox{act}(\{ \mbox{police} \}, \vertices{V}) + 
 \mbox{act}(\vertices{V}, \{ \mbox{police} \})$,
 $\func{fas} = \mbox{act}(\{ \mbox{fascists} \}, \vertices{V}) +
 \mbox{act}(\vertices{V}, \{ \mbox{fascists} \})$ and
 $all = \mbox{act}(\vertices{V}, \vertices{V})$. 
They are presented in Figure~\ref{viol}. Comparing the intensity
charts of police and fascists activity with overall activity we see
that most of the violent activities in the first two years 1919 and 1920
were related to the police. In the next two years (1921 and 1922) they were taken
over by the fascists.

\begin{figure}[!]
 \begin{center}
   \includegraphics[width=75mm]{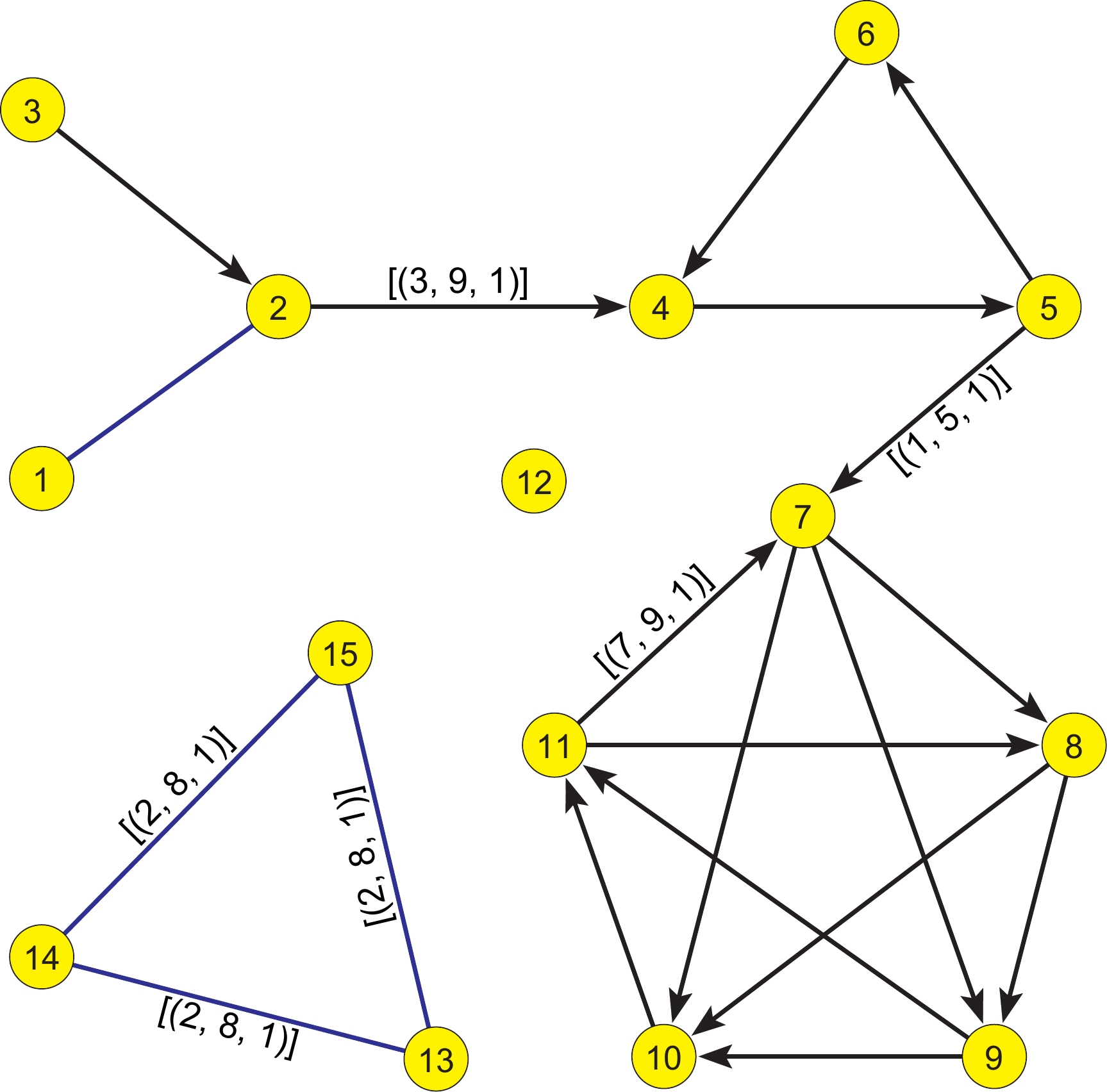}
  \caption{First example network. All unlabeled links have a value of $[(1,9,1)]$.\label{graph}}
 \end{center}
\end{figure}

\section{Temporal degrees \label{degrees}}

For an ordinary graph with a (binary) adjacency matrix $\mathbf{A}$ we can
compute the corresponding indegree, $\mathbf{i}$, and outdegree, $\mathbf{o}$, vectors using (over the combinatorial semiring) the relations
\[ \mathbf{i} = \mathbf{e} \bullet \mathbf{A} \qquad \mbox{and} \qquad
   \mathbf{o} =  \mathbf{A} \bullet \mathbf{e} \] 
where $\mathbf{e}$ is a column vector of size $n = |\vertices{V}|$ with all
its entries equal to 1. The same holds for temporal networks. In this case
the vector $\mathbf{e}$ contains as values the temporal unit  $\mathbf{1} = [(0,\infty,1)]$.

For a temporal network presented in Figure~\ref{graph} the corresponding
temporal indegrees and outdegrees are given in Table~\ref{iod}.
For example, the node 5 has in the time interval $[1,5)$ outdegree 2. Because
the arc $(5,7)$ disappears at the time point 5 the outdegree of the node 5
diminishes to 1 in the interval $[5,9)$.

\begin{table}
\caption{Temporal indegrees and outdegrees for the first example network.\label{iod}}
\begin{center}
{\renewcommand{\baselinestretch}{0.8}\small
\begin{verbatim}
     indegrees	            outdegrees
 1 : [(1, 9, 1)]       1 : [(1, 9, 1)]        
 2 : [(1, 9, 2)]       2 : [(1, 3, 1),         
 3 : []                     (3, 9, 2)]        
 4 : [(1, 3, 1),       3 : [(1, 9, 1)]
      (3, 9, 2)]       4 : [(1, 9, 1)]
 5 : [(1, 9, 1)]       5 : [(1, 5, 2),         
 6 : [(1, 9, 1)]            (5, 9, 1)]        
 7 : [(1, 5, 1),       6 : [(1, 9, 1)]
      (7, 9, 1)]       7 : [(1, 9, 3)]
 8 : [(1, 9, 2)]       8 : [(1, 9, 2)]        
 9 : [(1, 9, 2)]       9 : [(1, 9, 2)]        
10 : [(1, 9, 3)]      10 : [(1, 9, 1)]        
11 : [(1, 9, 2)]      11 : [(1, 7, 1),         
12 : []                     (7, 9, 2)]        
13 : [(2, 8, 2)]      12 : []        
14 : [(2, 8, 2)]      13 : [(2, 8, 2)]        
15 : [(2, 8, 2)]      14 : [(2, 8, 2)]        
                      15 : [(2, 8, 2)]
\end{verbatim}
}
\end{center}
\end{table}

We will use the simple temporal network from Figure~\ref{graph} also for the
illustration of some other algorithms because it allows the users to manually
check the presented results.

\section{Temporal co-occurrence networks}

Let the binary matrix $\mathbf{A}=[a_{ep}]$ describe a two-mode network on the set of
events $E$ and the set of of participants $P$:
\[  a_{ep} = \left\{\begin{array}{ll} 
                1 & p \mbox{ participated in the event } e \\
                0 & \mbox{otherwise}
             \end{array}\right. \]
The function $d: E \to \Time$ assigns to each event $e$ the date $d(e)$
when it happened. $\Time = [\func{first}, \func{last}]$. Using these data we can construct two
temporal affiliation matrices:
\begin{itemize}
\item \textbf{instantaneous} $\mathbf{Ai}=[ai_{ep}]$, where
\[  ai_{ep} = \left\{\begin{array}{ll} 
                [(d(e),d(e)+1,1)] & a_{ep} = 1 \\
                \lbrack\ \rbrack & \mbox{otherwise}
             \end{array}\right. \]
\item \textbf{cumulative} $\mathbf{Ac}=[ac_{ep}]$, where            
\[  ac_{ep} = \left\{\begin{array}{ll} 
                [(d(e),last+1,1)] & a_{ep} = 1 \\
                \lbrack\ \rbrack  & \mbox{otherwise}
             \end{array}\right. \]
\end{itemize}
Using the multiplication of temporal matrices over the combinatorial semiring we
get the corresponding instantaneous and cumulative co-occurrence matrices
\[  \mathbf{Ci} = \mathbf{Ai}^T \cdot \mathbf{Ai} \qquad \mbox{and}
    \qquad \mathbf{Cc} = \mathbf{Ac}^T \cdot \mathbf{Ac} \]             
A typical example of such a matrix is the papers authorship matrix where $E$ is the
set of papers, $P$ is the set of authors and $d$ is the publication year \citep{bibnet}.

\begin{table}
\caption{Temporal collaboration.\label{tec}}
\begin{center}
{\renewcommand{\baselinestretch}{0.8}\small
\begin{verbatim}
     ci[IDI/B,HCL/B]    cc[IDI/B,HCL/B]
 1 : (2003, 2004, 1)    (2003, 2004, 1)
 2 : (2004, 2005, 2)    (2004, 2005, 3)
 3 : (2005, 2006, 3)    (2005, 2006, 6)
 4 : (2006, 2007, 2)    (2006, 2007, 8)
 5 : (2007, 2008, 1)    (2007, 2008, 9)
 6 : (2008, 2009, 7)    (2008, 2009, 16)
 7 : (2009, 2010, 6)    (2009, 2010, 22)
 8 : (2010, 2011, 7)    (2010, 2011, 29)
 9 : (2011, 2013, 18)   (2011, 2012, 47)
10 :                    (2012, 2013, 65)

     ci[HCL/B,HCL/B]    cc[HCL/B,HCL/B]
 1 : (1997, 1998, 2)    (1997, 1998, 2)
 2 : (1998, 1999, 5)    (1998, 1999, 7)
 3 : (1999, 2000, 8)    (1999, 2000, 15)
 4 : (2000, 2001, 7)    (2000, 2001, 22)
 5 : (2001, 2002, 5)    (2001, 2002, 27)
 6 : (2002, 2003, 6)    (2002, 2003, 33)
 7 : (2003, 2004, 14)   (2003, 2004, 47)
 8 : (2004, 2005, 20)   (2004, 2005, 67)
 9 : (2005, 2006, 10)   (2005, 2006, 77)
10 : (2006, 2007, 14)   (2006, 2007, 91)
11 : (2007, 2008, 20)   (2007, 2008, 111)
12 : (2008, 2009, 28)   (2008, 2009, 139)
13 : (2009, 2010, 56)   (2009, 2010, 195)
14 : (2010, 2011, 78)   (2010, 2011, 273)
15 : (2011, 2012, 84)   (2011, 2012, 357)
16 : (2012, 2013, 112)  (2012, 2013, 469)
\end{verbatim}
}
\end{center}
\end{table}

The triple $(s,f,v)$ in a temporal quantity $ci_{pq}$ tells that in the time interval
$[s,f)$ there were $v$ events in which both $p$ and $q$ took part.

The triple $(s,f,v)$ in a temporal quantity $cc_{pq}$ tells that in the time interval
$[s,f)$ there were in total $v$ accumulated events in which both $p$ and $q$ took part.

The diagonal matrix entries  $ci_{pp}$ and  $cc_{pp}$ contain the temporal quantities
counting the number of events in the time intervals in which the participant
$p$ took part.

For example, in a data set on the stem cell research during 1997--2012 in Spain collected by
Gisela Cantos-Mateos \citep{stem} we get from the basic two-mode network, where $E$ is the
set of papers and $P$ is the set of institutions, for selected two institutions
(HCL/B $=$ University Hospital Cl\'{\i}nic de Barcelona, Barcelona and 
IDI/B $=$ Institut d'Investigacions Biom\`{e}diques August Pi i Sunyer, Barcelona)
the collaboration temporal quantities presented in Table~\ref{tec}.

The first column in the table contains the yearly collaboration (co-authorship) data
and the second column contains the cumulative collaboration data.
Let's read the table:\\
$ci[\texttt{IDI/B},\texttt{HCL/B}](2005, 2006) = 3$ --- in the year 2005 researchers from both institutions published 3 joint papers;\\
$ci[\texttt{IDI/B},\texttt{HCL/B}](2011, 2013) = 18$ --- in the years 2011 and 2012 researchers from both institutions published 18 joint papers each year;\\
$ci[\texttt{HCL/B},\texttt{HCL/B}](2010, 2011) = 78$ --- in the year 2010 researchers from the institution \texttt{HCL/B} published 78 papers;\\
$cc[\texttt{IDI/B},\texttt{HCL/B}](2008, 2009) = 16$ --- till the year 2008 (included) researchers from both institutions published 16 joint papers.

Note that the violence network from Section~\ref{activ} is essentially a
co-occurrence network that could be obtained from the more primitive
instantaneous two-mode network about violent actions reported in journal articles and the involved political actors. 

\section{Clustering coefficients}
Let us assume that the network $\Net$ is based on a simple directed graph $\network{G} = (\vertices{V},\edges{A})$
without loops. From a simple undirected graph we obtain the corresponding simple
directed graph by replacing each edge with a pair of opposite arcs. In such a 
graph the \keyw{clustering coefficient}, $C(v)$, of the node
$v$  is defined as the proportion between the number of realized arcs among the node's neighbors and the number
of all possible arcs among the node's neighbors $N(v)$, that is
\[ C(v) = \frac{|\edges{A}(N(v))|}{k(k-1)} \]
where $k$ is the number of neighbors of the node $v$. For a node $v$ without neighbors or with a single neighbor we set $C(v)=0$.

The clustering coefficient measures a local density of the node's neighborhood. A problem with its applications
in network analysis is that the identified densest neighborhoods are mostly very small. For this reason
we provided in Pajek the \keyw{corrected clustering coefficient}, $C'(v)$,
\[ C'(v) = \frac{|\edges{A}(N(v))|}{\Delta(k-1)} \]
where $\Delta$ is the maximum number of neighbors in the network.

To count the number of  realized arcs among the node's neighbors we use the observation that each arc forms a triangle
with links from its end-nodes to the node $v$; and that the number of triangles in a simple undirected graph
can be obtained as the diagonal value in the third power of the graph matrix
(over the combinatorial semiring). 

\begin{figure}[h!]
 \begin{center}
   \large
   \begin{tabular}{ccccc}
    \includegraphics[width=20mm]{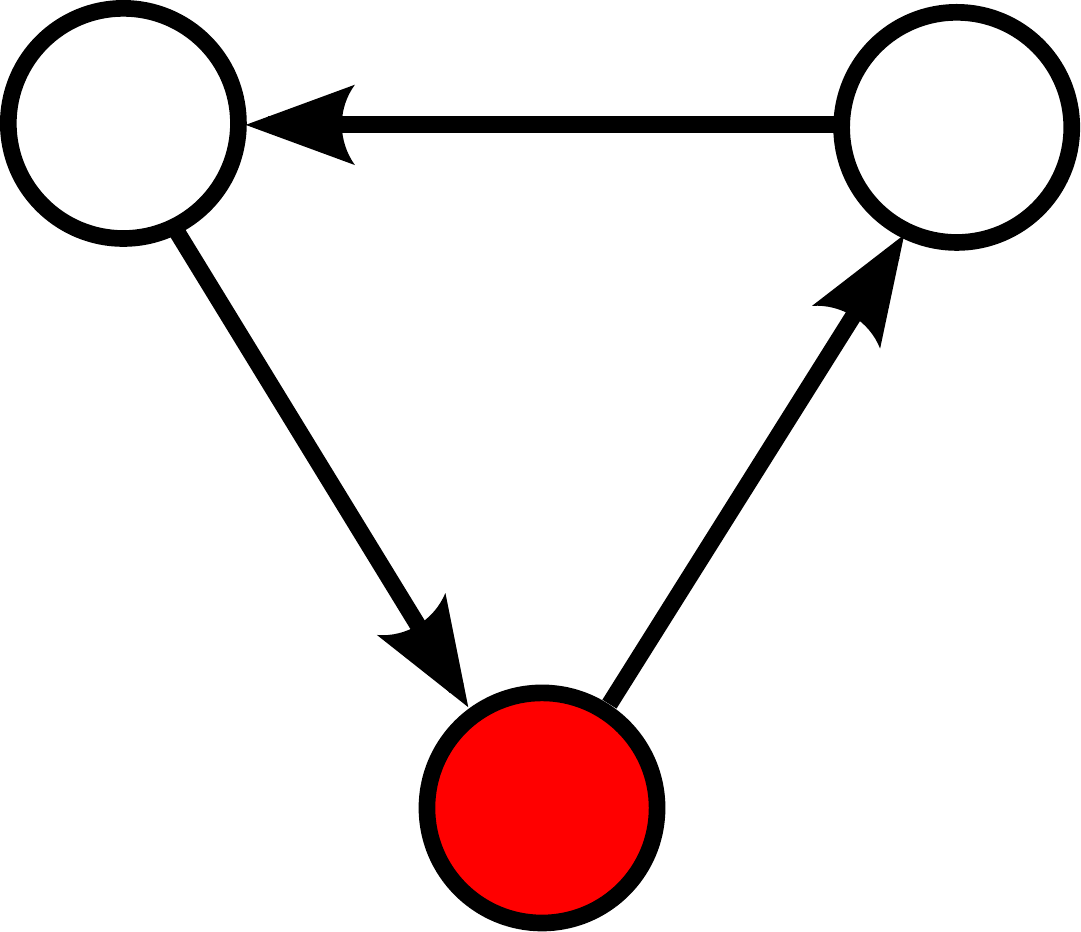} &\raisebox{7mm}{${\mathbf{AAA} \atop \mathbf{TTT}}$}& \qquad\qquad &
    \includegraphics[width=20mm]{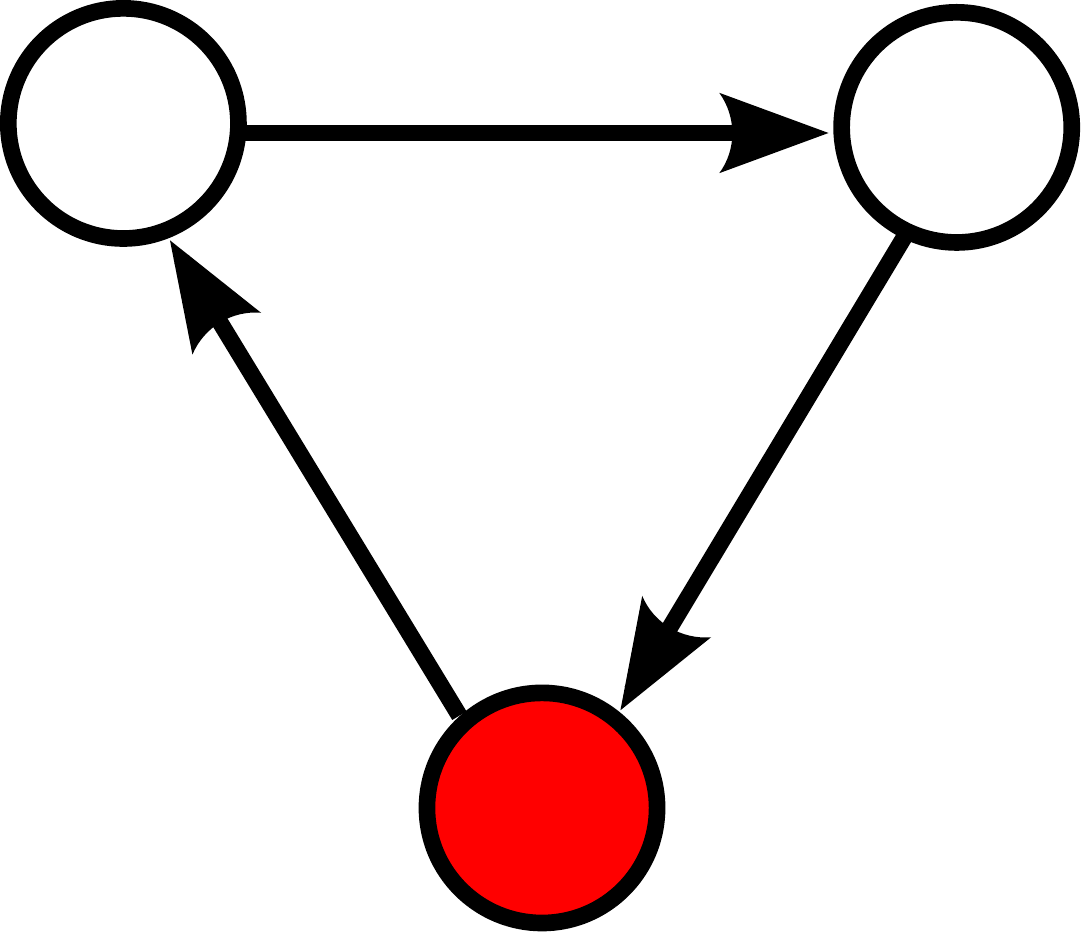} &\raisebox{7mm}{${\mathbf{TTT} \atop \mathbf{AAA}}$} \\[2mm]
    \includegraphics[width=20mm]{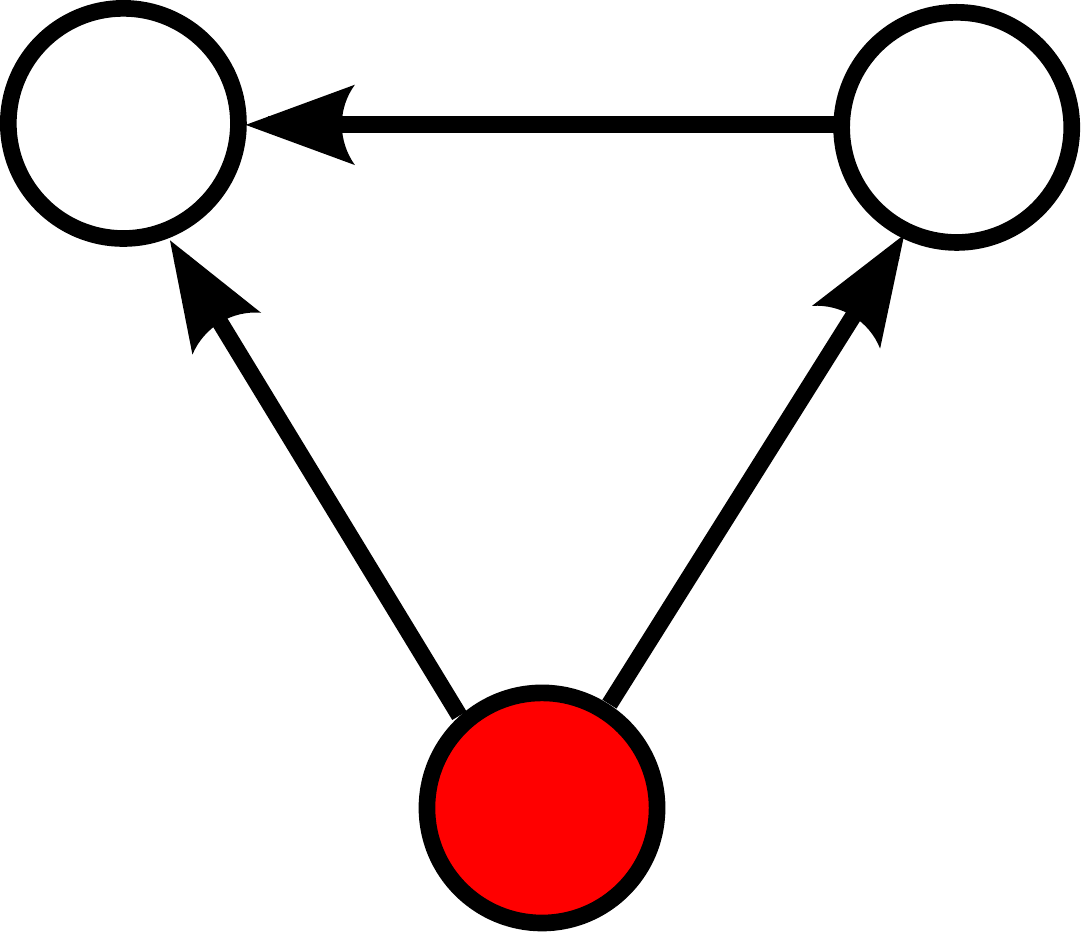} &\raisebox{7mm}{${\mathbf{AAT} \atop \mathbf{ATT}}$}& \qquad\qquad &
    \includegraphics[width=20mm]{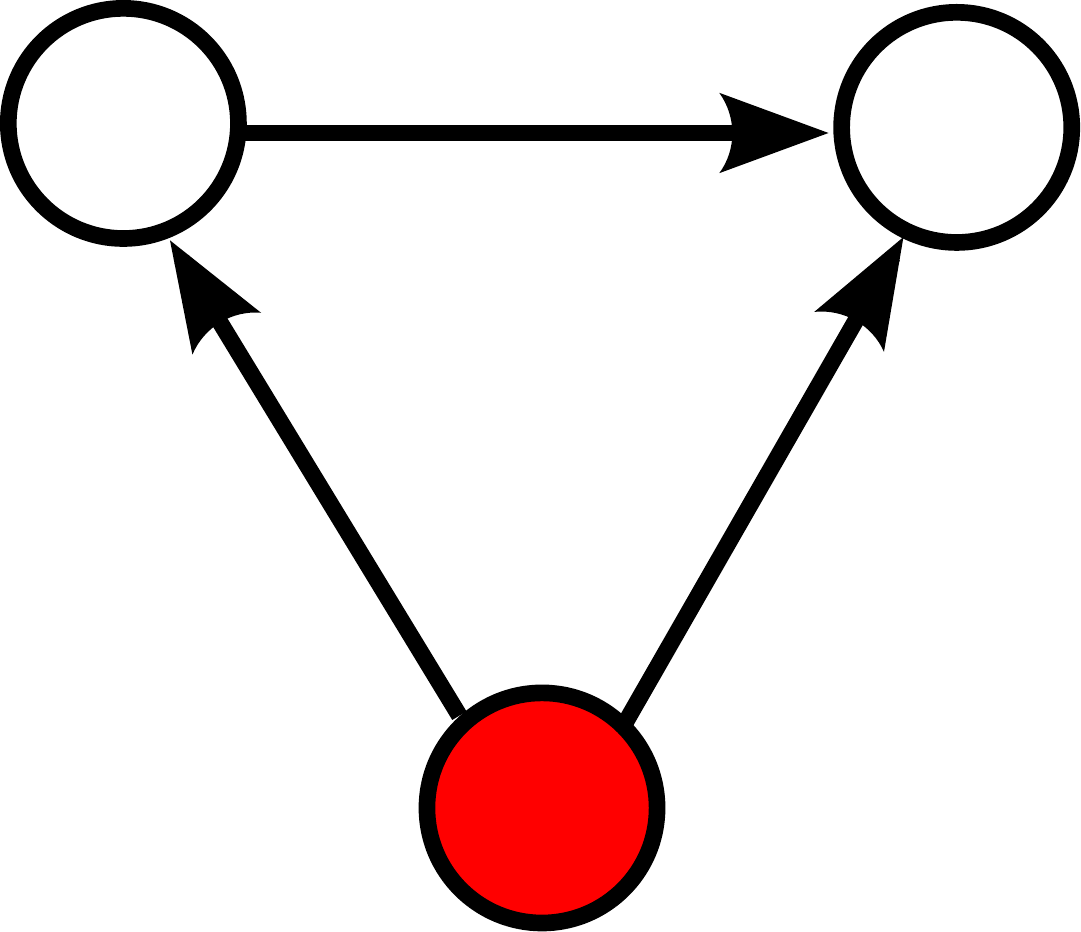} &\raisebox{7mm}{${\mathbf{ATT} \atop \mathbf{AAT}}$} \\[2mm]
    \includegraphics[width=20mm]{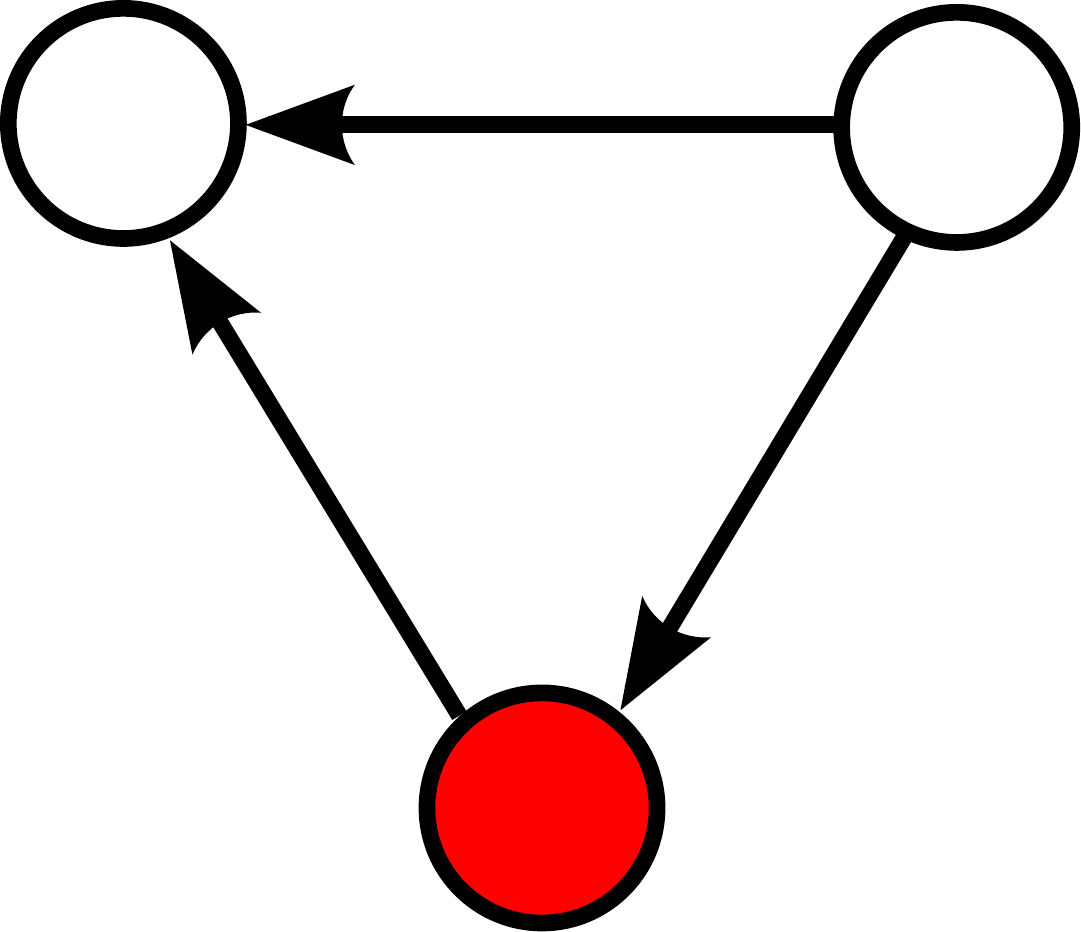} &\raisebox{7mm}{${\mathbf{TAT} \atop \mathbf{ATA}}$}& \qquad\qquad &
    \includegraphics[width=20mm]{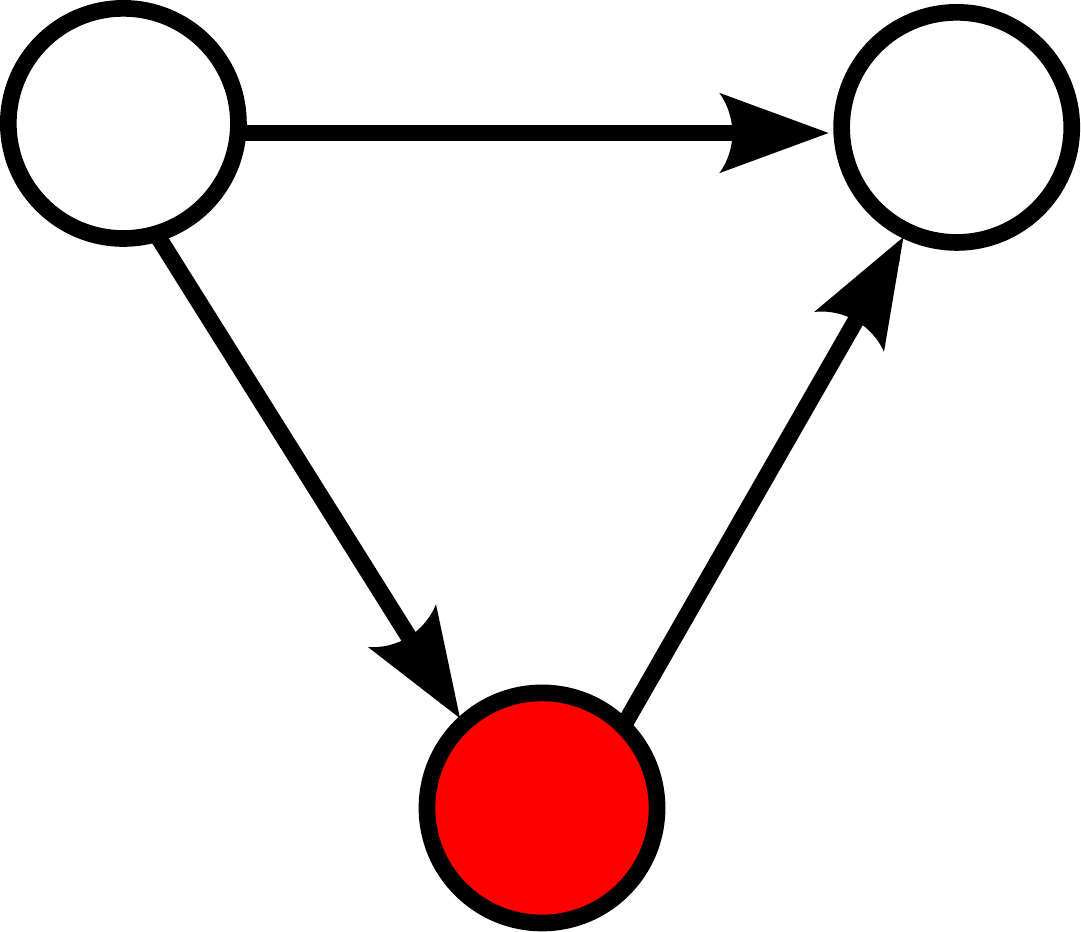} &\raisebox{7mm}{${\mathbf{ATA} \atop \mathbf{TAT}}$} \\[2mm]
    \includegraphics[width=20mm]{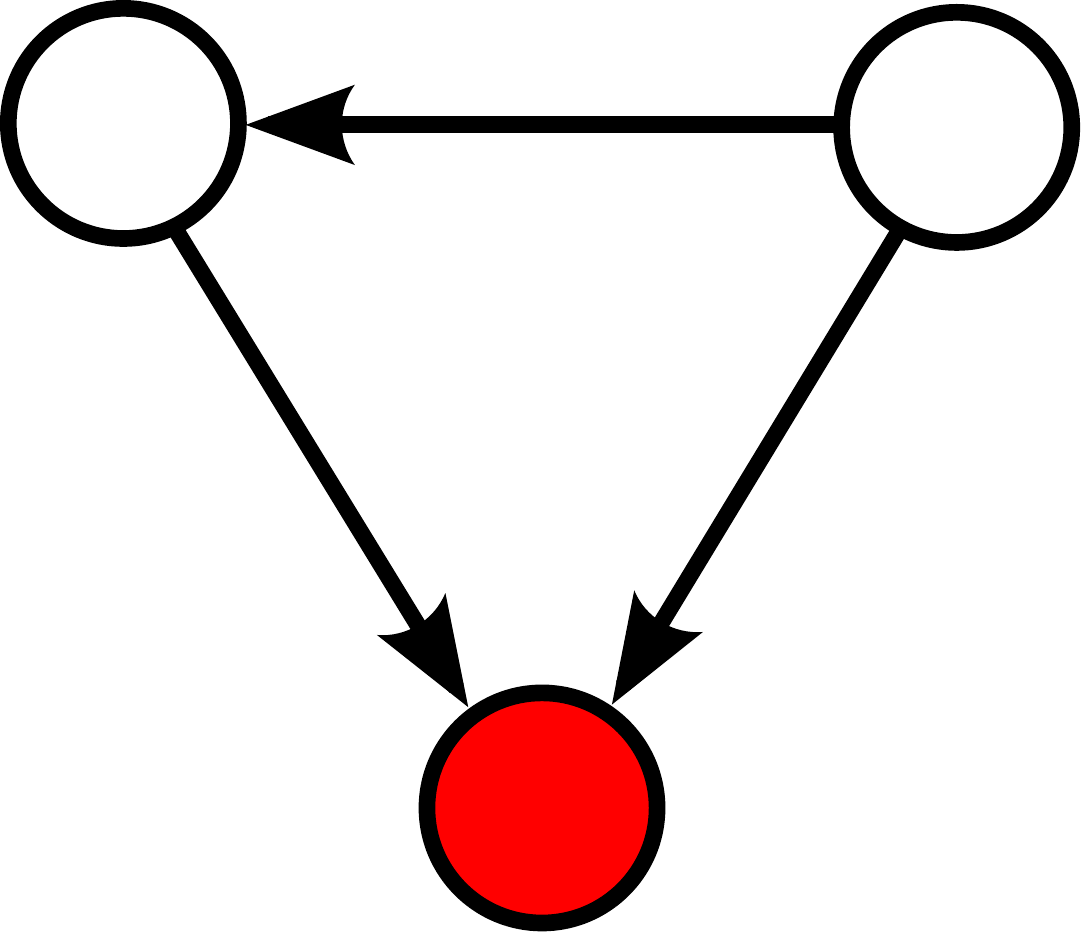} &\raisebox{7mm}{${\mathbf{TAA} \atop \mathbf{TTA}}$}& \qquad\qquad &
    \includegraphics[width=20mm]{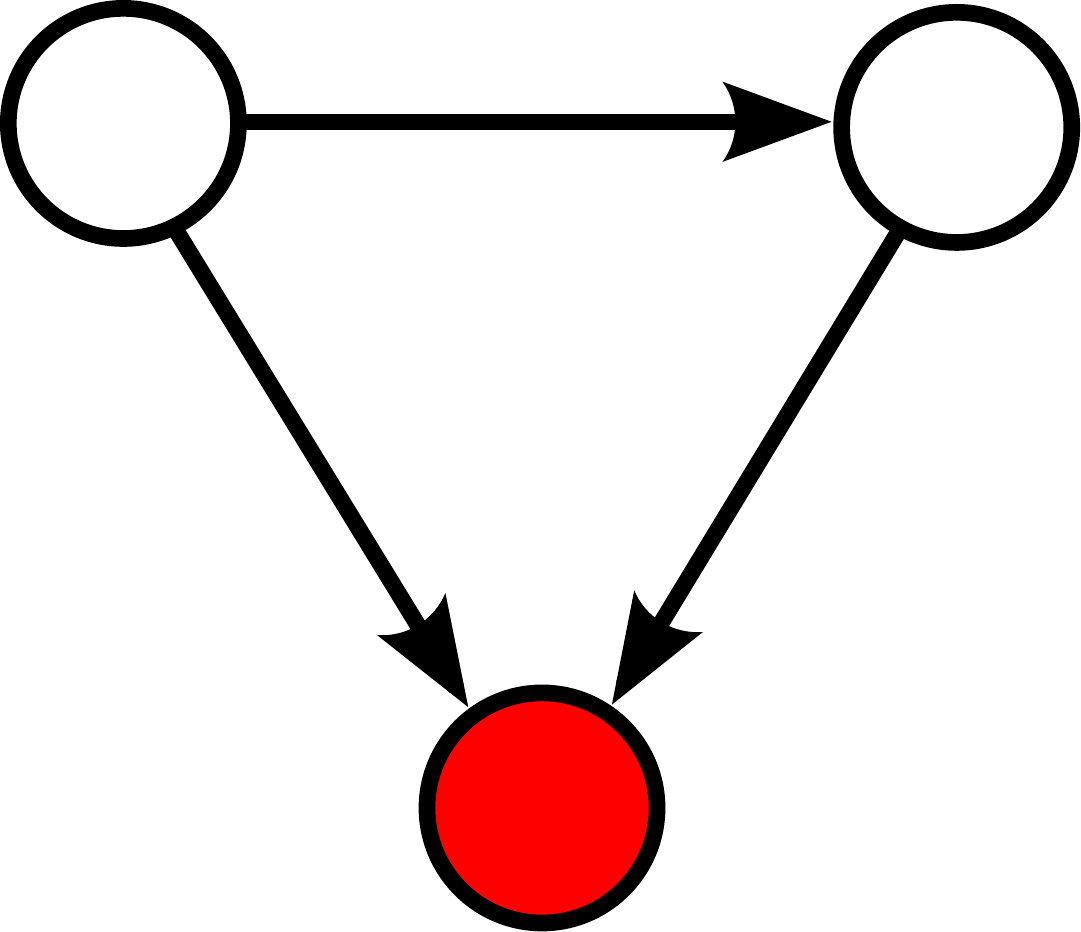} &\raisebox{7mm}{${\mathbf{TTA} \atop \mathbf{TAA}}$} 
   \end{tabular}
   \normalsize
  \caption{Counting triangles.\label{trik}}
 \end{center}
\end{figure}

For simple directed graphs the counting of triangles is slightly more complicated.
Let us denote $\mathbf{T} = \mathbf{A}^T$ and $\mathbf{S} = \mathbf{A}+\mathbf{T}$. From Figure~\ref{trik}
we see that each triangle (determined with a link opposite to the dark node) appears
exactly once in 
\[ \mathbf{AAA} + \mathbf{AAT} + \mathbf{TAT} + \mathbf{TAA} = \mathbf{AAS} + \mathbf{TAS} = \mathbf{SAS} . \] 
This gives us a simple way to count the triangles which is used in Algorithm~\ref{CCoef}.
The function $\func{nRows}(\mathbf{A})$ returns the size (number of rows) of matrix $\mathbf{A}$.
The function $\func{VecConst}(n,v)$ constructs a vector of size $n$ filled with the value $v$.
The function $\func{MatBin}(\mathbf{A})$ transforms all values in the triples in the matrix $\mathbf{A}$ to 1.
The function $\func{MatSetDiag}(\mathbf{A},c)$ sets all the diagonal entries of the matrix  $\mathbf{A}$ to the value $c$.
The function $\func{MatSym}(\mathbf{A})$ makes the transformation $\mathbf{S} = \mathbf{A}\oplus\mathbf{T}$. 
Functions $\func{VecSum}$ and $\func{VecProd}$ implement a component wise composition of temporal
vectors: $\func{VecSum}(a,b) = [a_i \oplus b_i,\ i = 1, \ldots, n]$ and 
$\func{VecProd}(a,b) = [a_i \odot b_i,\ i = 1, \ldots, n]$. Similarly 
$\func{VecInv}(a) = [\func{invert}(a_i),\ i = 1, \ldots, n]$ in the combinatorial semiring;
where $\func{invert}(a) = [ (s,f,1/v) \mbox{ \bf for } (s,f,v) \in a ]$. 
The function $\func{MatProd}(\mathbf{A},\mathbf{B})$ determines the product  $\mathbf{A \odot B}$.
Since we need only the diagonal values of the matrix $\mathbf{SAS}$ we applied a special
function $\func{MatProdDiag}(\mathbf{A},\mathbf{B})$ that determines 
only the diagonal vector of the product  $\mathbf{A \odot B}$.
Afterward, to get the clustering coefficient, we have to normalize the obtained
counts. The number of neighbors of the node $v$ is determined as its degree in the
corresponding undirected temporal skeleton graph (in which an edge $e=(v:u)$ exists iff there
is at least one arc between the nodes $v$ and $u$). The maximum number of neighbors 
$\Delta$ can be considered either for a selected time point ($\func{type}=2$) or for 
the complete time window ($\func{type}=3$).
Note that to determine the temporal $\Delta$ we used summing of temporal
degrees over the \keyw{maxmin} semiring $(\RR,\max,\min,-\infty,\infty)$.

The time complexity of Algorithm~\ref{CCoef} is $O(n^3 \cdot L)$.

In Table~\ref{cct} and Table~\ref{ccct} the ordinary and the corrected clustering
coefficients are presented for the example network from Figure~\ref{graph} and
its undirected skeleton.

\begin{algorithm}
\caption{Clustering coefficients.\label{CCoef}}
\begin{algorithmic}[1]

\Function{\func{clusCoef}}{$A,\func{type}=1$}\\
 \#{ \func{type} $= 1$ - standard CC} \\
 \#{ \func{type} $= 2$ - corrected CC / temporal degMax}\\
 \#{ \func{type} $= 3$ - corrected CC / overall degMax}
   \State $\func{SetSemiring}(\func{combinatorial})$ 
   \State $n \gets \func{nRows}(A)$; $ve \gets \func{VecConst}(n,[(0,\infty,-1)])$
   \State $B \gets \func{MatSetDiag}(\func{MatBin}(A),\mathbf{0})$
   \State $S \gets \func{MatBin}(\func{MatSym}(B))$
   \State $\func{deg} \gets \func{MatVecMulR}(S,\func{VecConst}(n,\mathbf{1}))$
   \If{$\func{type} = 1$}
      \State $\func{fac} \gets \func{VecProd}(\func{deg},\func{VecSum}(\func{deg},ve))$
   \Else
      \State $\func{SetSemiring}(\func{maxmin})$; $\delta \gets \mathbf{0}$
      \For{$d \in \func{deg}$} $\delta \gets \func{sum}(\delta,d)$ \EndFor
      \If{$\func{type} = 3$}
         \State $\Delta \gets \max([v \mbox{ \bf for } (s,f,v) \in \delta])$
         \State $\delta \gets [(0,\infty,\Delta)]$
      \EndIf
      \State $\func{SetSemiring}(\func{combinatorial})$
      \State $\func{degm} \gets \func{VecSum}(\func{deg},ve)$; $\func{fac} \gets \mathbf{0}$
      \For{$d \in \func{degm}$} $\func{fac}.\func{append}(\func{prod}(\delta,d))$ \EndFor
   \EndIf      
   \State $\func{tri} \gets \func{MatProdDiag}(\func{MatProd}(S,B),S)$
   \State \Return $\func{VecProd}(\func{VecInv}(\func{fac}),\func{tri})$
\EndFunction      

\end{algorithmic}
\end{algorithm}

\begin{table}
\caption{Clustering coefficients for the first example network.\label{cct}}
\begin{center}
{\renewcommand{\baselinestretch}{0.8}\small
\begin{verbatim}
 1 : []                           
 2 : []                           
 3 : []                           
 4 : [(1, 3, 0.5), (3, 9, 0.1667)]
 5 : [(1, 5, 0.1667), (5, 9, 0.5)]
 6 : [(1, 9, 0.5)]                
 7 : [(1, 5, 0.25), (5, 9, 0.5)]                                    
 8 : [(1, 7, 0.4167), (7, 9, 0.5)]
 9 : [(1, 7, 0.4167), (7, 9, 0.5)]
10 : [(1, 7, 0.4167), (7, 9, 0.5)]
11 : [(1, 9, 0.5)]                
12 : []                           
13 : [(2, 8, 1.0)]                
14 : [(2, 8, 1.0)]                
15 : [(2, 8, 1.0)]          
\end{verbatim}
}
\end{center}
\end{table}

\begin{table}
\caption{Corrected clustering coefficients for the skeleton of the first example network.\label{ccct}}
\begin{center}
{\renewcommand{\baselinestretch}{0.8}\small
\begin{verbatim}
 1 : []
 2 : []
 3 : []
 4 : [(1, 3, 0.5), (3, 9, 0.25)]
 5 : [(1, 5, 0.25), (5, 9, 0.5)]
 6 : [(1, 9, 0.5)]
 7 : [(1, 5, 0.5), (5, 7, 0.75),
      (7, 9, 1.0)]
 8 : [(1, 7, 0.8333), (7, 9, 1.0)]
 9 : [(1, 7, 0.8333), (7, 9, 1.0)]
10 : [(1, 7, 0.8333), (7, 9, 1.0)]
11 : [(1, 7, 0.75), (7, 9, 1.0)]
12 : []
13 : [(2, 8, 0.5)]
14 : [(2, 8, 0.5)]
15 : [(2, 8, 0.5)]
\end{verbatim}
}
\end{center}
\end{table}

\section{Closures in temporal networks}

When the basic semiring $(A,\oplus,\odot,0,1)$ is \keyw{closed} -- an unary \keyw{closure} operation $\star$
with the property 
\[ a^\star = 1 \oplus a\odot a^\star = 1 \oplus a^\star \odot a , \qquad 
\mbox{for all } a \in A \]
is defined in it -- this property can be extended also to the corresponding
matrix semiring. When it exists, a standard closure is obtained as
\[ a^\star = \bigoplus_{i=0}^\infty a^i . \]
In some semirings different closures can exist.
For computing the matrix closure
we can apply the Fletcher's algorithm \citep{closure}.
The entry $c_{uv}$ in the matrix $\mathbf{C} = \mathbf{A}^\star$ is equal to the sum of
values of all walks  from the node $u$ to the node $v$.
In most of the semirings, except the combinatorial, for which we are interested in determining the 
closures, also the \keyw{absorption law} holds
\[  1\oplus a=1, \qquad \mbox{for all } a \in A .\]
In these semirings $a^\star = 1$, for all $a \in A$, and therefore the
Fletcher's algorithm can be simplified and performed in place as
implemented in Algorithm~\ref{closeTQ}.

For a temporal quantity $a$ over a closed semiring it holds $T_{a^\star} = \Time$.

The time complexity of Algorithm~\ref{closeTQ} is $O(n^3 \cdot L)$.

\begin{algorithm}
\caption{Closure of a temporal matrix over an absorptive semiring.}
\label{closeTQ}
\begin{algorithmic}[1]

\Function{\func{MatClosure}}{$R,\func{strict}=\func{False}$}
   \State $n \gets \func{nRows}(R)$
   \State $C \gets R$
   \For{$k \in 1:n$}
      \For{$u \in 1:n$}
         \For{$v \in 1:n$}
            \State $C[u,v] \gets$
            \State \strut\qquad$\func{sum}(C[u,v], \func{prod}(C[u,k],C[k,v]))$
         \EndFor
      \EndFor
      \If{$\lnot \func{strict}$} $C[k,k] \gets \func{sum}(\mathbf{1},C[k,k])$ \EndIf
   \EndFor
   \State \Return $C$
\EndFunction      

\end{algorithmic}
\end{algorithm}

\section{Temporal node partitions}

In the previous sections, the nodes of temporal networks were considered as
being present all the time. We can describe the presence of nodes through time 
using a temporal binary (single valued) node partition $T : \vertices{V} \to A_{\scriptsize\cmdkey}(\Time)$,
\[ T(u) = ( (s_i, f_i, 1) )_{i=1}^k, \quad \mbox{for } u \in \vertices{V} \]
specifying that a node $u$ is present in time intervals $[s_i, f_i)$,
$i = 1, \ldots, k$.

The node partition $T_{Min}$ determined from the temporal network links by
\[ T_{Min}(u) = \bigcup_{l \in \edges{L}: u \in \mbox{\scriptsize ext}(l)} \func{binary}(a_l) , \]
for $ u \in \vertices{V}$, is the smallest temporal partition of nodes that
satisfies the consistency condition from Section~\ref{desc}.
The term $\mbox{ext}(l)$ denotes the set of endnodes of the link $l$, $a_l$ is
the temporal quantity assigned to the link $l$, and the function $\func{binary}$
sets all values in a given temporal quantity to 1.
In the library TQ the partition $T_{Min}$ can be computed using the function $\func{minTime}$.

A temporal node partition $q$ can also be used to extract a corresponding
subnetwork from the given temporal network described with a matrix $\mathbf{A}$.
The subnetwork contains only the nodes active in the partition $q$ and the active links
satisfying the consistency condition with respect to $q$.

To formalize the described procedure we first define the procedure
$\func{extract}(p,a) = b$, where $p$ is a binary temporal quantity and $a$ is a temporal quantity, as
\[ b(t) =  \left\{\begin{array}{ll} 
                a(t) & t \in T_p \cap T_a \\
                \cmdkey         & \mbox{otherwise}
             \end{array}\right. .
\]   
Let $\mathbf{B}$ be a temporal matrix describing the links of the subnetwork
determined by the partition $q$.
Its entries for $l(u,v) \in \edges{L}$ are determined by
\[ b_l = \func{extract}(q(u) \cap q(v),a_l) .  \]
In TQ this operation is implemented as a procedure 
$\func{MatExtract}(\mathbf{q},\mathbf{A})$.

\section{Temporal reachability and weak and strong connectivity \label{conn}}

For a temporal network represented with the corresponding binary matrix $\mathbf{A}$
its transitive closure $\mathbf{A}^\star$ (over the reachability
semirings based on the semiring $(\{0,1\},\lor,\land,0,1)$) determines its
\keyw{reachability} relation matrix.
We obtain its \keyw{weak connectivity} temporal matrix $\mathbf{W}$ as
\[  \mathbf{W} = (\mathbf{A} \cup \mathbf{A}^T)^\star \] 
and its \keyw{strong connectivity} temporal matrix $\mathbf{S}$ as
\[  \mathbf{S} = \mathbf{A}^\star \cap (\mathbf{A}^\star)^T  . \]
The use of the strict transitive closure instead of a transitive closure in these
relations preserves the inactivity value $\mathbf{0}$ on the diagonal for all 
isolated nodes.

\subsection{Reachability degrees}

Let $\mathbf{R} = \overline{\mathbf{A}} = \mathbf{A} \odot \mathbf{A}^\star$ 
be the strict reachability relation of
a given network. Then the temporal vectors $\func{inReach} = \func{inDeg}(\mathbf{R})$ and $\func{outReach} = \func{outDeg}(\mathbf{R})$ 
contain temporal quantities counting the number of nodes: from which a given
node $v$ is reachable $( \func{inReach}[v] )$ / which are reachable
from the node $v$ $( \func{outReach}[v] )$. The results for our example network
are presented in Table~\ref{ioReach}. For example, 8 nodes 
$\{ 4,5,6,7,8,9,10,11 \}$ are reachable from  node 6 in the time interval $[1,5)$,
and  3 nodes $\{ 4,5,6 \}$ are reachable in the time interval $[5,9)$.

\begin{table}
\caption{Temporal input reachability degrees for the first example network.\label{ioReach}}
\begin{center}
{\renewcommand{\baselinestretch}{0.8}\small
\begin{verbatim}
 1 : [(1, 9, 3)]                         
 2 : [(1, 9, 3)]                         
 3 : []                                  
 4 : [(1, 3, 3), (3, 9, 6)]              
 5 : [(1, 3, 3), (3, 9, 6)]              
 6 : [(1, 3, 3), (3, 9, 6)]              
 7 : [(1, 3, 3), (3, 5, 6), (7, 9, 5)]   
 8 : [(1, 3, 8), (3, 5, 11), (5, 9, 5)]  
 9 : [(1, 3, 8), (3, 5, 11), (5, 9, 5)]  
10 : [(1, 3, 8), (3, 5, 11), (5, 9, 5)]  
11 : [(1, 3, 8), (3, 5, 11), (5, 9, 5)]  
12 : []                                  
13 : [(2, 8, 3)]                         
14 : [(2, 8, 3)]                         
15 : [(2, 8, 3)]                  
\end{verbatim}
}
\end{center}
\end{table}

\subsection{Temporal weak connectivity}

The function $\func{weakConnMat}(\mathbf{A})$ for a given temporal network  
matrix $\mathbf{A}$ determines the corresponding temporal weak connectivity
matrix $\mathbf{W}$. Every time slice $\Net(t)$, $t \in \Time$, of the matrix 
 $\mathbf{W}$ is an
equivalence relation that can be compactly described with the corresponding
partition.

To transform the temporal equivalence matrix $\mathbf{E}$  into the corresponding
temporal partition $\mathbf{p}$ we use the fact that on a given time interval
equivalent (in our case weakly connected) nodes get the same value on this interval
in the product of the matrix $\mathbf{E}$ with a vector computed over the
combinatorial semiring $(\NN,+,\cdot,0,1)$. We take for the
vector values randomly shuffled integers from the interval $1:n$. With a very high
probability the values belonging to different equivalence classes are different.
This is implemented as a procedure $\func{eqMat2Part}(\mathbf{E})$ (see Algorithm~\ref{eqparTQ}).
Maybe in the future implementations we shall add a loop with the check of the
injectivity of this mapping. The classes of the obtained temporal partition are finally
renumbered with consecutive numbers using the function $renumPart(p)$
(see Algorithm~\ref{renumTQ}). The variable $C$ in the description of the function
$renumPart$ is a dictionary (data structure). 

For our first example network we obtain the temporal weak partition presented on the
left hand side of Table~\ref{weakC}.

\begin{algorithm}
\caption{Transform temporal equivalence relation into partition.}
\label{eqparTQ}
\begin{algorithmic}[1]

\Function{\func{eqMat2Part}}{$E$}
   \State $\func{SetSemiring}(\func{combinatorial})$
   \State $v \gets \func{shuffle}([ [(0,\infty,i+1)] \mbox{ \bf for } i \in 1:\func{nRows}(E)])$
   \State $p \gets \func{MatVecMulR}(E,v)$
   \State \Return $\func{renumPart}(p)$ 
\EndFunction      
\end{algorithmic}
\end{algorithm}

\begin{algorithm}
\caption{Renumber the classes of a partition.}
\label{renumTQ}
\begin{algorithmic}[1]

\Function{\func{renumPart}}{$p$}
   \State $C \gets \{\ \}$; $q = [\ ]$
   \For{$a \in p$}
      \State $r \gets [\ ]$
      \For{$(s_a,f_a,c_a) \in a$}
         \If{$c_a \notin C$} $C[c_a] \gets 1+\func{length}(C)$ \EndIf
         \State $r.\func{append}((s_a,f_a,C[c_a]))$
      \EndFor
      \State $q.\func{append}(r)$
    \EndFor
    \State \Return $q$
\EndFunction      
\end{algorithmic}
\end{algorithm}

\subsection{Temporal strong connectivity}

The procedure $\func{strongConnMat}(\mathbf{A})$ for a given temporal network  
matrix $\mathbf{A}$ determines the corresponding temporal strong connectivity
matrix $\mathbf{S}$. To determine the intersection of temporal network binary matrices 
$\mathbf{A}$ and $\mathbf{B}$ we use the function $\func{MatInter}(\mathbf{A},\mathbf{B})$. 
Again, to get the strong connectivity partition we have to apply the 
function $\func{eqMat2Part}$ to the strong connectivity matrix.

The time complexity of algorithms for temporal weak and strong connectivity partitions is $O(n^3 \cdot L)$.

For our first example network we obtain the temporal strong partition presented on 
the right hand side of Table~\ref{weakC}. In the library TQ both matrices and 
partitions are based on the strict transitive closure.

\begin{table}
\caption{Temporal weak and strong connectivity partitions for the first example network.\label{weakC}}
\begin{center}
{\renewcommand{\baselinestretch}{0.8}\small
\begin{verbatim}
Weak partition                             
 1 : [(1, 3, 1), (3, 5, 2), (5, 9, 3)]     
 2 : [(1, 3, 1), (3, 5, 2), (5, 9, 3)]     
 3 : [(1, 3, 1), (3, 5, 2), (5, 9, 3)]     
 4 : [(1, 3, 4), (3, 5, 2), (5, 9, 3)]     
 5 : [(1, 3, 4), (3, 5, 2), (5, 9, 3)]     
 6 : [(1, 3, 4), (3, 5, 2), (5, 9, 3)]     
 7 : [(1, 3, 4), (3, 5, 2), (5, 9, 5)]     
 8 : [(1, 3, 4), (3, 5, 2), (5, 9, 5)]     
 9 : [(1, 3, 4), (3, 5, 2), (5, 9, 5)]     
10 : [(1, 3, 4), (3, 5, 2), (5, 9, 5)]     
11 : [(1, 3, 4), (3, 5, 2), (5, 9, 5)]     
12 : []                                    
13 : [(2, 8, 6)]                           
14 : [(2, 8, 6)]                           
15 : [(2, 8, 6)]   

Strong partition
 1 : [(1, 9, 1)]
 2 : [(1, 9, 1)]
 3 : []
 4 : [(1, 9, 2)]
 5 : [(1, 9, 2)]
 6 : [(1, 9, 2)]
 7 : [(7, 9, 3)]
 8 : [(1, 7, 4), (7, 9, 3)]
 9 : [(1, 7, 4), (7, 9, 3)]
10 : [(1, 7, 4), (7, 9, 3)]
11 : [(1, 7, 4), (7, 9, 3)]
12 : []
13 : [(2, 8, 5)]
14 : [(2, 8, 5)]
15 : [(2, 8, 5)]               
\end{verbatim}
}
\end{center}
\end{table}

\section{Temporal closeness and betweenness}

Closeness and betweenness are among the traditional social network analysis
indices measuring the importance of nodes \citep{cent}. They are somehow problematic
when applied to non (strongly) connected graphs. In this section we will not
consider these questions. We will only show how to compute them for
non-problematic temporal graphs.
 
\subsection{Temporal closeness}

The \keyw{output closeness} of the node $v$ is defined as
\[ ocl(v) = \frac{n-1}{\displaystyle\sum_{u \in \vertices{V}\setminus \{v\}} d_{vu}} .\]

To determine the closeness we first need to compute the matrix  $\mathbf{D} = [d_{uv}]$ of geodetic distances  $d_{uv}$ between the nodes $u$ and $v$. It can be obtained as a closure of the network matrix $\mathbf{A}$ over the
\keyw{shortest paths} semiring $(\overline{\RR_0^+},\min,+,\infty,0)$.
Note that the values in the matrix $\mathbf{A}$ can be any nonnegative real numbers.

\begin{figure}[!]
 \begin{center}
   \includegraphics[width=75mm]{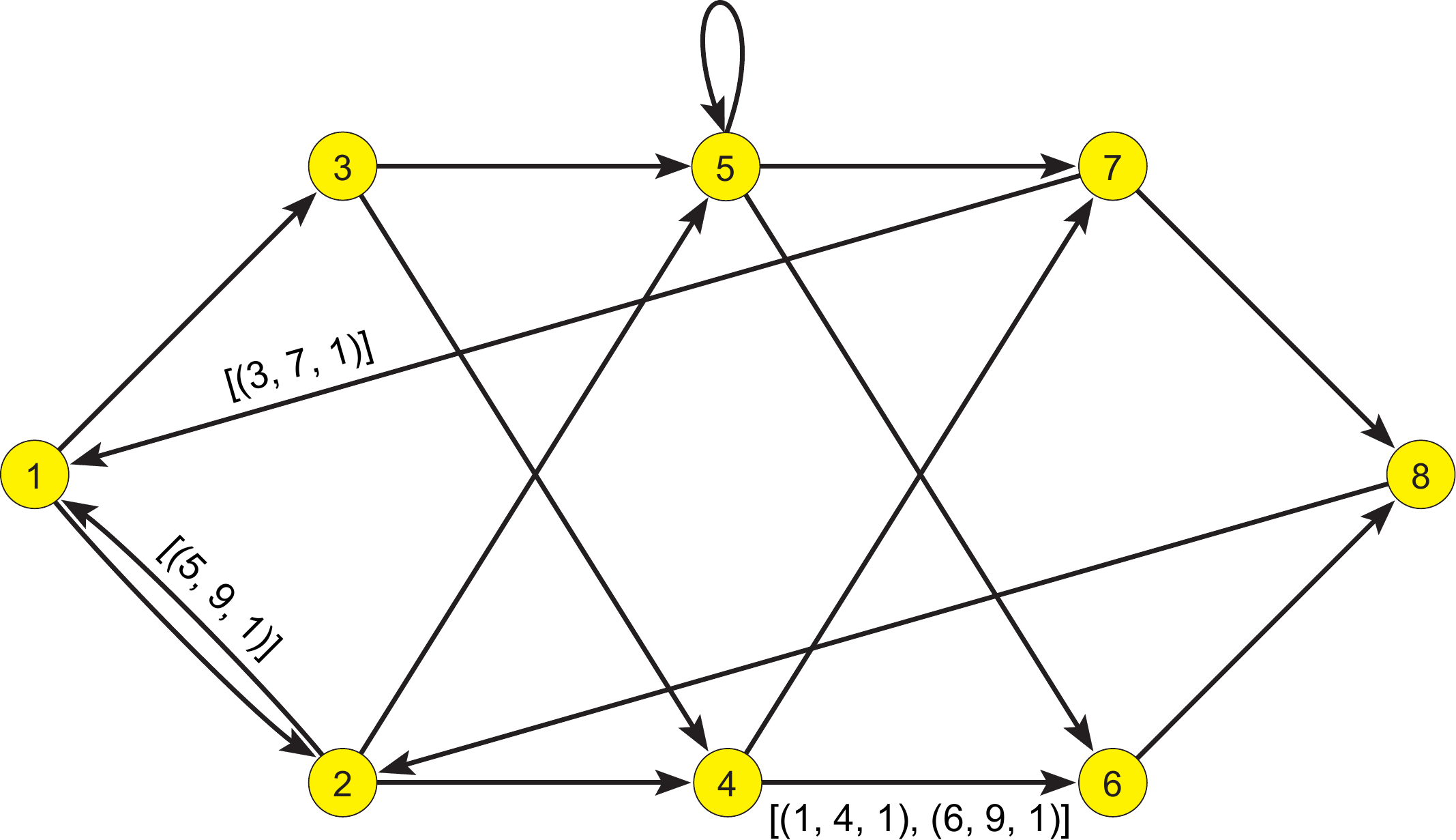}
  \caption{Second example network. All unlabeled arcs have the value $[(1,9,1)]$.\label{graph2}}
 \end{center}
\end{figure}

In Figure~\ref{graph2} we present our second example temporal network which
is an extended version of the example given in Figure~3 from \cite{semi}.

Because a complete strict closure matrix $\mathbf{D}$ is too large
to be listed we present only some of its selected entries: 

{\renewcommand{\baselinestretch}{0.8}\small
\begin{verbatim}
D[3,1] = [(3, 7, 3), (7, 9, 5)]
D[4,6] = [(1, 4, 1), (4, 6, 5), (6, 9, 1)]
D[6,3] = [(3, 5, 6), (5, 9, 4)]
D[7,6] = [(1, 9, 4)]
\end{verbatim}
}

To compute the vector of closeness coefficients of nodes we have to sum
the temporal distances to other nodes over the combinatorial semiring.
See  Algorithm~\ref{closTQ}.
While summing we replace gaps (inactivity intervals inside ${\cal T}$) with time intervals
with the value infinity, using the procedure $\func{fillGaps}$.
The time complexity of Algorithm~\ref{closTQ} is $O(n^3 \cdot L)$.

The temporal closeness coefficients for our second example network are 
given in Table~\ref{cloC}.

\begin{algorithm}
\caption{Temporal closeness.\label{closTQ}}
\begin{algorithmic}[1]

\Function{\func{closeness}}{$A,\func{type}=2$}\\
\# {\func{type}: 1 - output, 2 - all, 3 - input}
   \State $s \gets \func{startTime}(A)$; $f \gets \func{finishTime}(A)$
   \State $n \gets \func{nRows}(A)$
   \State $\func{SetSemiring}({path})$
   \State $D \gets \func{MatClosure}(A,\func{strict}=True)$
   \State $\func{SetSemiring}(\func{combinatorial})$
   \State $k \gets (2-|\func{type}-2|)\cdot (n-1)$; $\func{fac} \gets [(0,\infty,k)]$ 
   \For{$v \in 1:n$}
      \State $d \gets \mathbf{0}$
      \For{$u \in 1:n$} 
         \If{$u \ne v$}
            \If{$\func{type}<3$}
            \State $d \gets \func{sum}(d,\func{fillGaps}(D[v,u],s,f))$ \EndIf
            \If{$\func{type}>1$} 
            \State $d \gets \func{sum}(d,\func{fillGaps}(D[u,v],s,f))$ \EndIf
         \EndIf
      \EndFor   
      \State $cl[v] \gets \func{prod}(\func{fac},\func{invert}(d))$
   \EndFor   
   \State \Return $cl$
\EndFunction      
\end{algorithmic}
\end{algorithm}

\begin{table}[!]
\caption{Output closeness for the second example network.\label{cloC}}
\begin{center}
{\renewcommand{\baselinestretch}{0.8}\small
\begin{verbatim}
1 : [(1, 9, 0.4375)]
2 : [(1, 3, 0.0000), (3, 5, 0.4375), 
     (5, 9, 0.5833)]
3 : [(1, 3, 0.0000), (3, 7, 0.4375),
     (7, 9, 0.3889)]
4 : [(1, 3, 0.0000), (3, 4, 0.4375),
     (4, 6, 0.3500), (6, 7, 0.4375),
     (7, 9, 0.3500)]
5 : [(1, 3, 0.0000), (3, 7, 0.4375),
     (7, 9, 0.3500)]
6 : [(1, 3, 0.0000), (3, 5, 0.2917),
     (5, 9, 0.3500)]
7 : [(1, 3, 0.0000), (3, 7, 0.4375),
     (7, 9, 0.3500)]
8 : [(1, 3, 0.0000), (3, 5, 0.3500),
     (5, 9, 0.4375)]
\end{verbatim}
}
\end{center}
\end{table}

\subsection{Temporal betweenness}

The \keyw{betweenness} of a node $v$ is defined as
\[  b(v) = \frac{1}{(n-1)(n-2)} \sum_{u,w \in \vertices{V} \atop |\{v,u,w\}| = 3}
    \frac{n_{u,w}(v)}{n_{u,w}}  \]
where $n_{u,w}$ is the number of $u$-$w$ geodesics (shortest paths) and $n_{u,w}(v)$
is the number of $u$-$w$ geodesics passing through the node $v$.

Suppose that we know the matrix
\[ \mathbf{C} = \lbrack ( d_{u,v}, n_{u,v} ) \rbrack \]
where $d_{u,v}$ is the length of $u$-$v$ geodesics. Then it is also easy
to determine the quantity $n_{u,w}(v)$:
\[  n_{u,w}(v) = \left\{ \begin{array}{ll}
          n_{u,v} \cdot n_{v,w} \qquad & d_{u,v} + d_{v,w} = d_{u,w} \\
          0  & \mbox{\rm otherwise}
   \end{array} \right.  . \]
This gives the following scheme of procedure for computing the nontemporal betweenness coefficients 
$\mathbf{b}$ \medskip

\begin{algorithmic}[1]

  \State compute $\mathbf{C}$
  \For{$v \in \vertices{V}$}
     \State $r \gets 0$
     \For{$u \in \vertices{V}, w \in \vertices{V}$}
        \If{$n[u,w]\ne 0 \land |\{v,u,w\}| = 3 \land \hspace*{15mm} d[u,w]=d[u,v]+d[v,w]$}
           \State  $r \gets r + n[u,v]\cdot n[v,w]/n[u,w]$
        \EndIf
     \EndFor
     \State  $b[v] \gets r / ((n-1)\cdot (n-2))$
  \EndFor      
\end{algorithmic}
\medskip  

In \cite{semi} it is shown that the matrix $\mathbf{C}$ can be obtained by computing
the closure of the network matrix over the \keyw{geodetic semiring} 
$(\overline{\NN}^2, \oplus, \odot,$ $ (\infty,0), (0,1))$, where
$\overline{\NN} = \NN\cup \{ \infty \}$ and
we define {\em addition} $\oplus$ with
\[  (a,i) \oplus (b,j) = ( \min(a,b), \left\{ \begin{array}{ll}
        i & a < b \\ i+j \quad & a = b \\ j & a > b
   \end{array} \right. )  \] 
and {\em multiplication} $\odot$ with:
\[  (a,i) \odot (b,j) = (a+b,i\cdot j) . \] 

To compute the geodetic closure we first transform the network temporal 
adjacency matrix $\mathbf{A}$  to a matrix  $\mbox{\bf G} = \lbrack (d,n)_{u,v} \rbrack$ which has for entries
pairs defined by
\[  (d,n)_{u,v}(t) = \left\{ \begin{array}{ll}
          (1,1) \quad  & \exists l \in \edges{L}: l(u,v) \land t \in T(l) \\
          \cmdkey       & \mbox{otherwise}
   \end{array} \right.  \] 
where $d$ is the length of a geodesic and $n$ is the number
of geodesics from $u$ to $v$. In temporal networks the distance $d$ and
the counter $n$ are temporal quantities.

The presented scheme adapted for computing the temporal betweenness vector is
implemented in TQ as the function $\func{betweenness}(A)$. First we compute its 
strict geodetic closure $\mathbf{C}$ over the geodetic semiring.
We present only some of its selected entries for our second example network: 

{\renewcommand{\baselinestretch}{0.8}\small
\begin{verbatim}   
C[1,7] = [(1, 9, (3, 4))]
C[2,2] = [(1, 3, (4, 4)), (3, 4, (4, 6)),
         (4, 5, (4, 5)), (5, 9, (2, 1))]
C[4,6] = [(1, 4, (1, 1)), (4, 6, (5, 3)), 
         (6, 9, (1, 1))]
C[5,5] = [(1, 9, (1, 1))]
C[6,3] = [(3, 5, (6, 2)), (5, 9, (4, 1))]
C[7,6] = [(1, 3, (4, 2)), (3, 4, (4, 6)), 
         (4, 6, (4, 3)), (6, 7, (4, 6)),
         (7, 9, (4, 2))]
\end{verbatim}
}

For example, the value $\mathbf{C}[4,6]$ reflects the facts that an arc
exists from node 4 to node 6 in time intervals $[1,4)$ and $[6,9)$; and
in the time interval $[4,6)$ they are connected with 3 geodesics of length 5:
$(4,7,8,2,5,6)$, $(4,7,1,3,5,6)$, $(4,7,1,2,5,6)$.

We continue and using the combinatorial semiring we compute the
temporal betweenness vector $\mathbf{b}$. The specificity of temporal
quantities $d[u,v]$ and $n[u,v]$ is considered in the auxiliary function 
$\func{between}$ that implements the temporal version of the statement
\medskip

{\bf if} $ d[u,w]=d[u,v]+d[v,w]$ {\bf then}\\
\hspace*{8mm} $r \gets r + n[u,v]\cdot n[v,w]/n[u,w]$\medskip

\noindent
from the basic betweenness algorithm. Again we apply the merging
scheme. The time complexity of the procedure $\func{betweenness}$ is
 $O(n^3 \cdot L)$.

The temporal betweenness coefficients for our second example network are
presented in Table~\ref{betwC}. 
\begin{table}
\caption{Betweenness for the second example network.\label{betwC}}
\begin{center}
{\renewcommand{\baselinestretch}{0.8}\small
\begin{verbatim}
1 : [(3, 4, 0.2500), (4, 6, 0.2754),
     (6, 7, 0.2500), (7, 9, 0.1429)]
2 : [(1, 3, 0.3452), (3, 4, 0.4048),
     (4, 6, 0.4187), (6, 7, 0.4048),
     (7, 9, 0.6071)]
3 : [(1, 3, 0.0595), (3, 4, 0.0952),
     (4, 6, 0.1052), (6, 7, 0.0952),
     (7, 9, 0.0595)]
4 : [(1, 3, 0.1667), (3, 4, 0.2500),
     (4, 5, 0.1762), (5, 6, 0.1048), 
     (6, 9, 0.1786)]
5 : [(1, 3, 0.1667), (3, 4, 0.2500),
     (4, 5, 0.3476), (5, 6, 0.2762),
     (6, 9, 0.1786)]
6 : [(1, 3, 0.1190), (3, 4, 0.0952),
     (4, 6, 0.0544), (6, 7, 0.0952),
     (7, 9, 0.1786)]
7 : [(1, 3, 0.1190), (3, 4, 0.4048),
     (4, 5, 0.4694), (5, 6, 0.3266),
     (6, 7, 0.2619), (7, 9, 0.1786)]
8 : [(1, 3, 0.3095), (3, 4, 0.2500),
     (4, 6, 0.2484), (6, 7, 0.2500),
     (7, 9, 0.5238)]
\end{verbatim}
}
\end{center}
\end{table}

\section{Temporal PathFinder}

The Pathfinder algorithm was proposed in the eighties  \citep{PF88,PF90} for the simplification of
weighted networks -- it removes from the network all links that do
not satisfy the (generalized) triangle inequality -- if for a weighted link there exists a shorter
path connecting its endnodes then the link is removed.
The basic idea of the Pathfinder algorithm is simple. It produces
a network $\mbox{PFnet}(\mathbf{W},r,q) = (\vertices{V},\edges{L}_{PF})$
determined by the following scheme of procedure \medskip

\begin{algorithmic}[1]
  \State compute $\mathbf{W}^{(q)}$; 
  \State $\edges{L}_{PF} \gets \emptyset$; 
  \For{$e(u,v) \in \edges{L}$} 
     \If{$\mathbf{W}^{(q)}[u,v] = \mathbf{W}[u,v]$}
        \State $\edges{L}_{PF} \gets \edges{L}_{PF} \cup \{ e \}$ 
     \EndIf
  \EndFor
\end{algorithmic}
\medskip

\noindent
where $\mathbf{W}$ is a network \emph{\textbf{dissimilarity}} matrix and 
$\mathbf{W}^{(q)} = \bigoplus_{i=1}^q \mathbf{W}^i =
(\mathbf{1} \oplus \mathbf{W})^q$ is the
matrix of the values of all walks of length at most $q$ computed over the 
\keyw{Pathfinder} semiring
$(\overline{\RR^+_0},\oplus,\Mw,\infty,0)$ with $a \Mw b = \sqrt[r]{a^r+b^r}$ and
$a \oplus b = \min(a,b)$. The value of $w_{uv}(q)$ in the matrix $\mathbf{W}^{(q)}$
is equal to the value of all walks of length at most $q$ from the node $u$ to the node $v$.

The scheme of Pathfinder is implemented as the function $\func{pathFinder}$. 
The temporal version of the statement

{\bf if} $\mathbf{W}^{(q)}[u,v] = \mathbf{W}[u,v]$ {\bf then}
     $\edges{L}_{PF} := \edges{L}_{PF} \cup \{ e \}$ 
     
\noindent
is implemented in the function $\func{PFcheck}$ using the merging scheme.

The function $\func{MatPower}(A,k)$ computes the $k$-th power of the matrix $\mathbf{A}$.

The time complexity of Algorithm~\ref{pfTQ}$+$\ref{pfcTQ} is $O(L \cdot n^3 \cdot \log q)$  \citep{PFc}.

\begin{algorithm}[h!]
\caption{Temporal PathFinder.\label{pfTQ}}
\begin{algorithmic}[1]

\Function{\func{pathFinder}}{$W,r=1,q=\infty$}
   \State $n \gets \func{nRows}(W)$; $\func{SetSemiring}(\func{pathfinder},r,q)$
   \If{$q>n$} $Z \gets \func{MatClosure}(W)$ \Else $\ Z \gets \func{MatPower}(\func{MatSetDiag}(W,\mathbf{1}),q)$ \EndIf
   \For{$u \in 1:n, v \in 1:n$}
      \State $PF[u,v] \gets \func{PFcheck}(W[u,v],Z[u,v])$
   \EndFor
   \State \Return $PF$
\EndFunction      
\end{algorithmic}
\end{algorithm}
            
\begin{algorithm}[h!]
\caption{Temporal PathFinder merge operation.\label{pfcTQ}}
\begin{algorithmic}[1]

\Function{\func{PFcheck}}{$a,b$}
   \If{$\func{length}(a) = 0$} \Return $a$ \EndIf
   \If{$\func{length}(b) = 0$} \Return $a$ \EndIf 
   \State $c \gets [\ ]$
   \State $(s_a,f_a,v_a) \gets \func{get}(a)$; $(s_b,f_b,v_b) \gets \func{get}(b)$
   \While{$(s_a<\infty) \lor (s_b<\infty)$}
      \If{$f_a \le s_b$} $(s_a,f_a,v_a) \gets \func{get}(a)$
      \ElsIf{$f_b \le s_a$} $(s_b,f_b,v_b) \gets \func{get}(b)$
      \Else
         \State $s_c \gets \max(s_a,s_b)$; $f_c \gets \min(f_a,f_b)$
         \If{$v_b = v_a$} $c.\func{append}((s_c,f_c,v_a))$ \EndIf
         \If{$f_c = f_a$} $(s_a,f_a,v_a) \gets \func{get}(a)$ \EndIf
         \If{$f_c = f_b$} $(s_b,f_b,v_b) \gets \func{get}(b)$ \EndIf
      \EndIf
   \EndWhile
   \State \Return $\func{standard}(c)$
\EndFunction      
\end{algorithmic}
\end{algorithm}

\begin{figure}[!]
 \begin{center}
   \includegraphics[width=75mm]{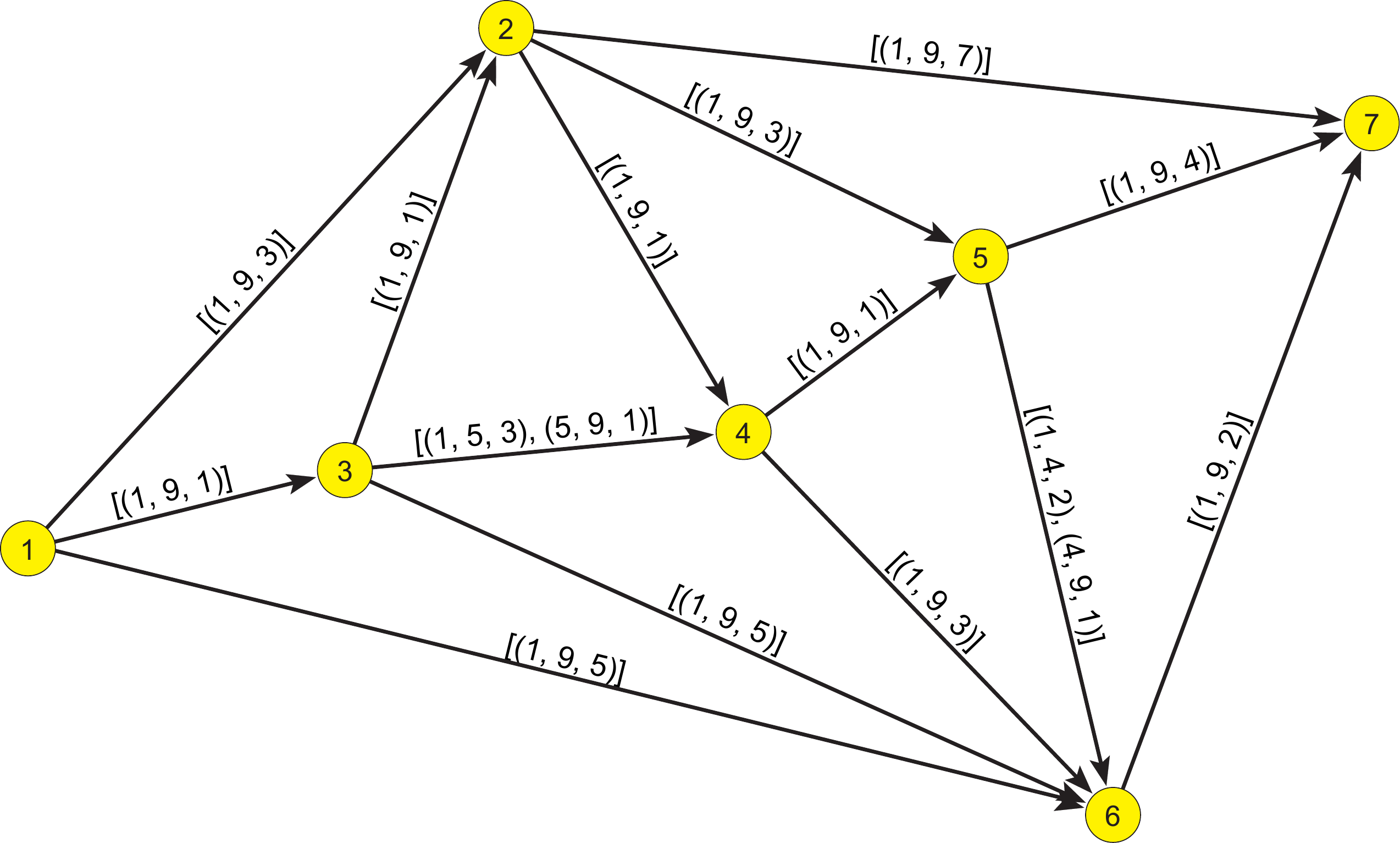} \\
   \includegraphics[width=75mm]{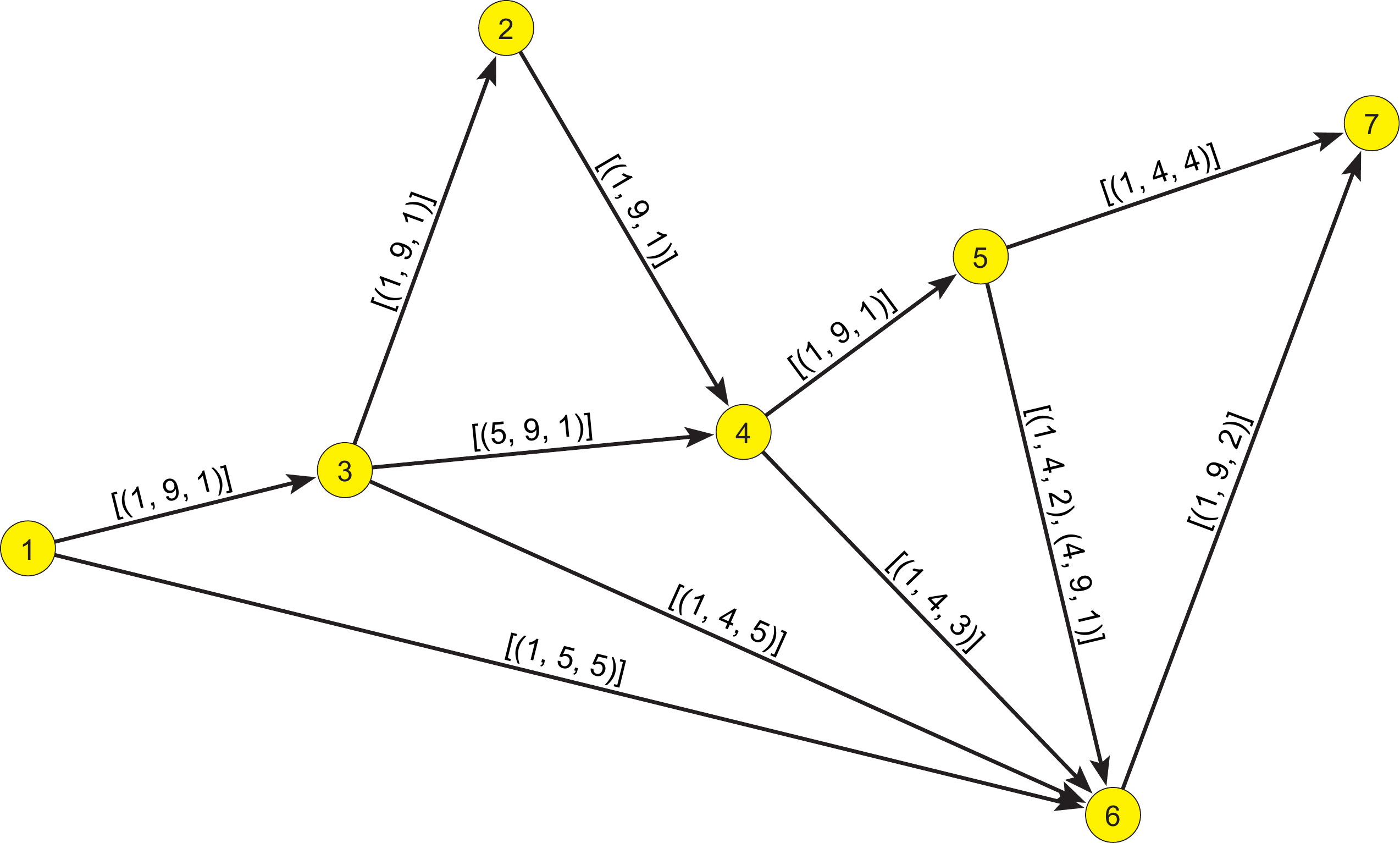}
 \caption{Pathfinder example.\label{PFex}}
 \end{center}
\end{figure}

The bottom network in Figure~\ref{PFex} presents the Pathfinder skeleton
$\mbox{PFnet}(\Net,1,\infty)$ of a network $\Net$ presented in the top part of the
same figure. Because $r=1$ a link $e$ is removed if there exists a path, connecting
its initial node to its terminal node, with the value (sum of link values) smaller than 
the value of the link $e$.
The arc $(1,2)$ is removed because $3 = v(1,2) > v(1,3)+v(3,2) = 2$. 
The arc $(1,6)$ is removed in the time interval $[5,9)$ because in this interval
 $5 = v(1,6) > v(1,3)+v(3,4)+v(4,5) +v(5,6)  = 4$.

\section{September 11th Reuters terror news}

The Reuters terror news network was obtained from the CRA (Centering Resonance Analysis) networks produced by
Steve Corman and Kevin Dooley at Arizona State University. 
The network is based on all the stories released during 66 consecutive days
by the news agency Reuters concerning the September 11 attack on the U.S.,
beginning at 9:00 AM EST 9/11/01.
The nodes of this network are important words (terms). There is an edge between
two words iff they appear in the same utterance (for details see the paper  \cite{CRA}). The
weight of an edge is its frequency. The network has $n = 13332$
nodes (different words in the news) and $m = 243447$ edges, 50859
with value larger than 1. There are no loops in the network.

The Reuters terror news network was used as a case network for the
Viszards visualization session on the Sunbelt XXII International
Sunbelt Social Network Conference, New Orleans, USA, 13-17. February 2002.

We transformed the Pajek version of the network into the Ianus format used in TQ.
To identify important terms we computed their aggregated frequencies
and extracted the subnetwork of the 50 most frequently used (during 66 days) nodes.
They are listed in  Table~\ref{terrF}. 

\begin{figure*}[!]
 \begin{center}
   \includegraphics[width=140mm,viewport=85 50 820 570,clip=]{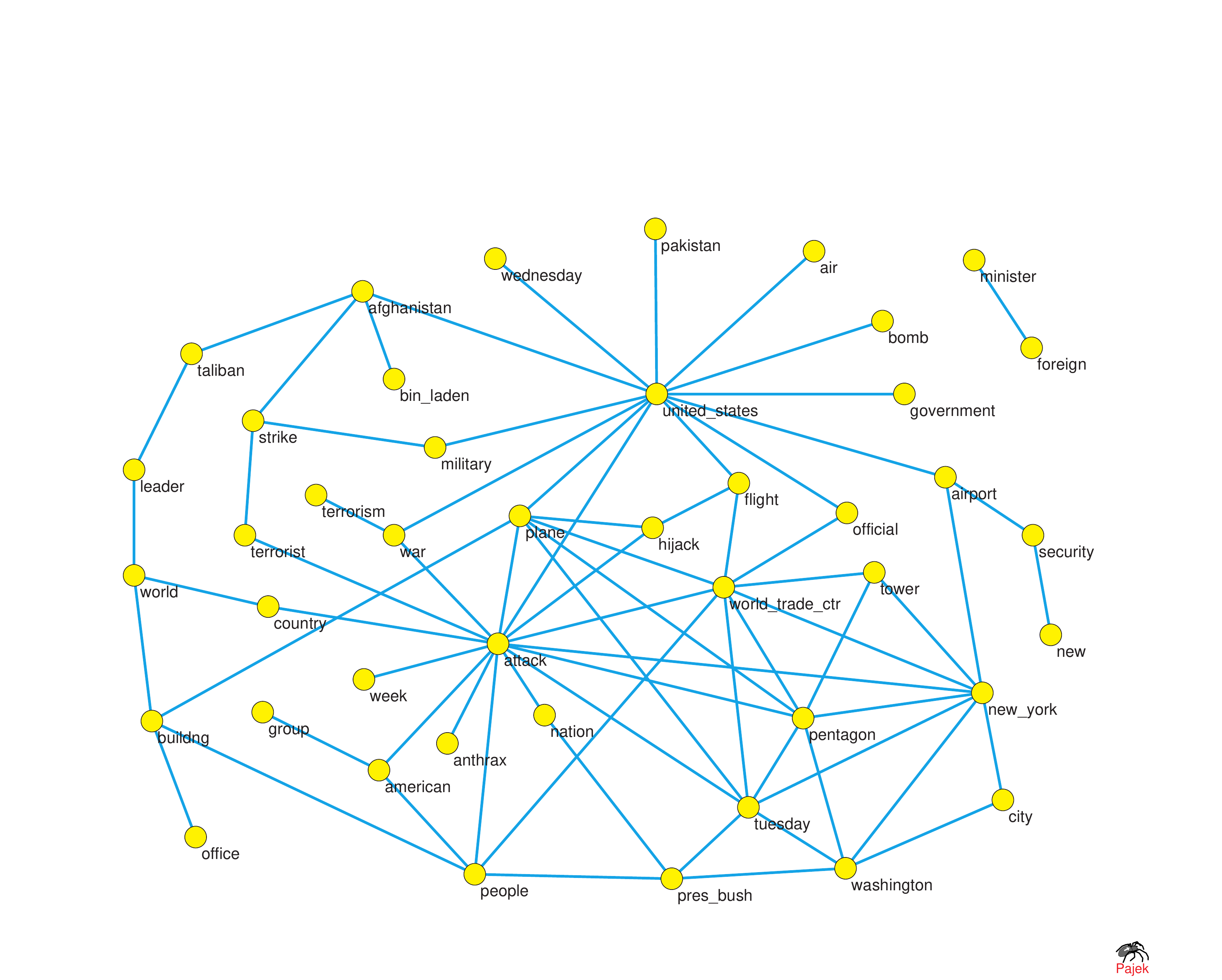}
 \caption{September 11th.\label{terrEx}}
 \end{center}
\end{figure*}

Trying to draw this subnetwork it turns out to be almost
a complete graph. To obtain something readable we removed all 
temporal edges with a value smaller than 10. The corresponding
underlying graph is presented in Figure~\ref{terrEx}. The isolated nodes
were removed.

\begin{figure}[!]
 \begin{center}
    \begin{tabular}{l}
    hijack :\\
   \includegraphics[width=70mm,viewport=80 165 780 535,clip=]{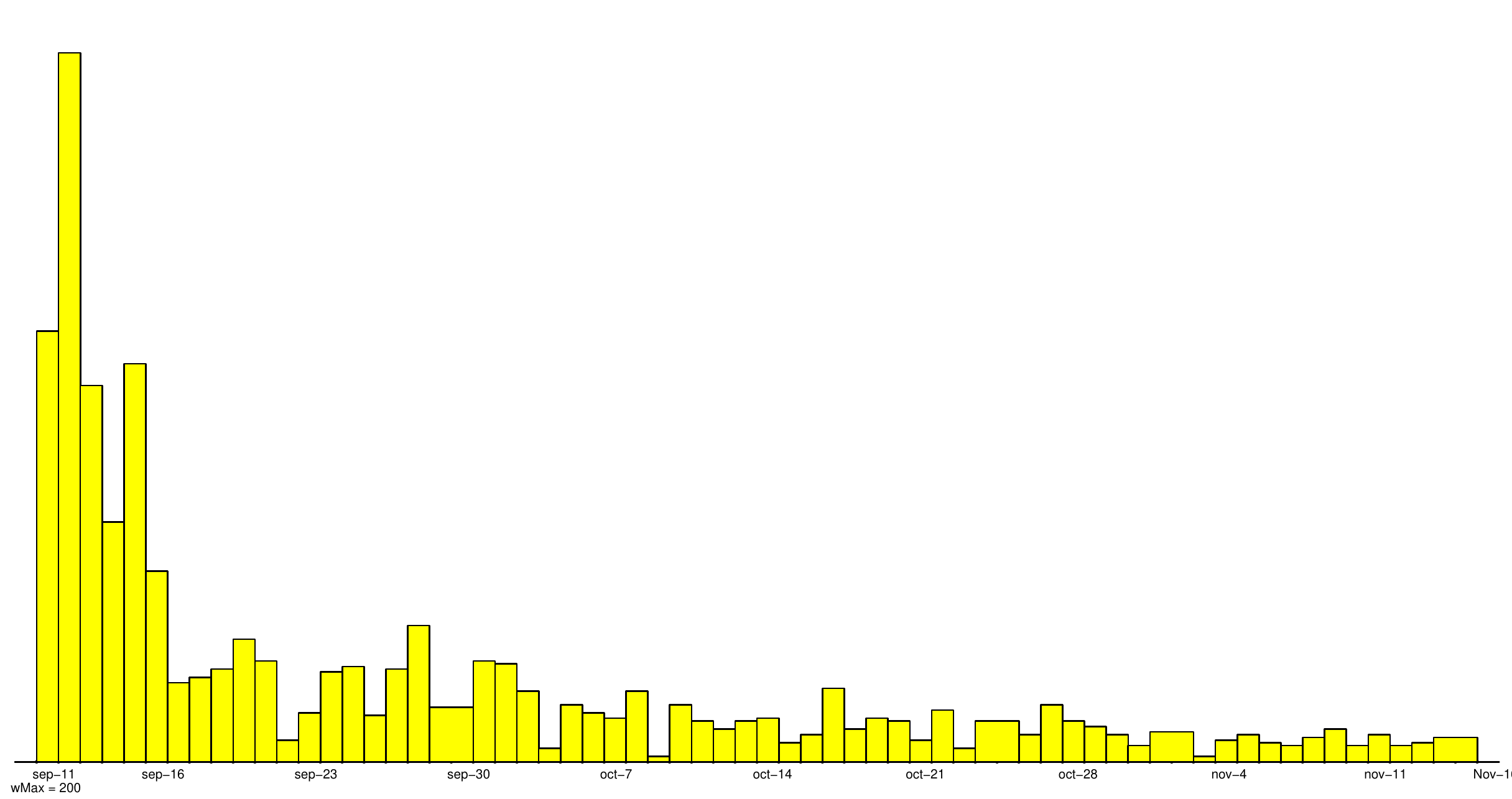}\\
   bin Laden :\\
   \includegraphics[width=70mm,viewport=80 165 780 455,clip=]{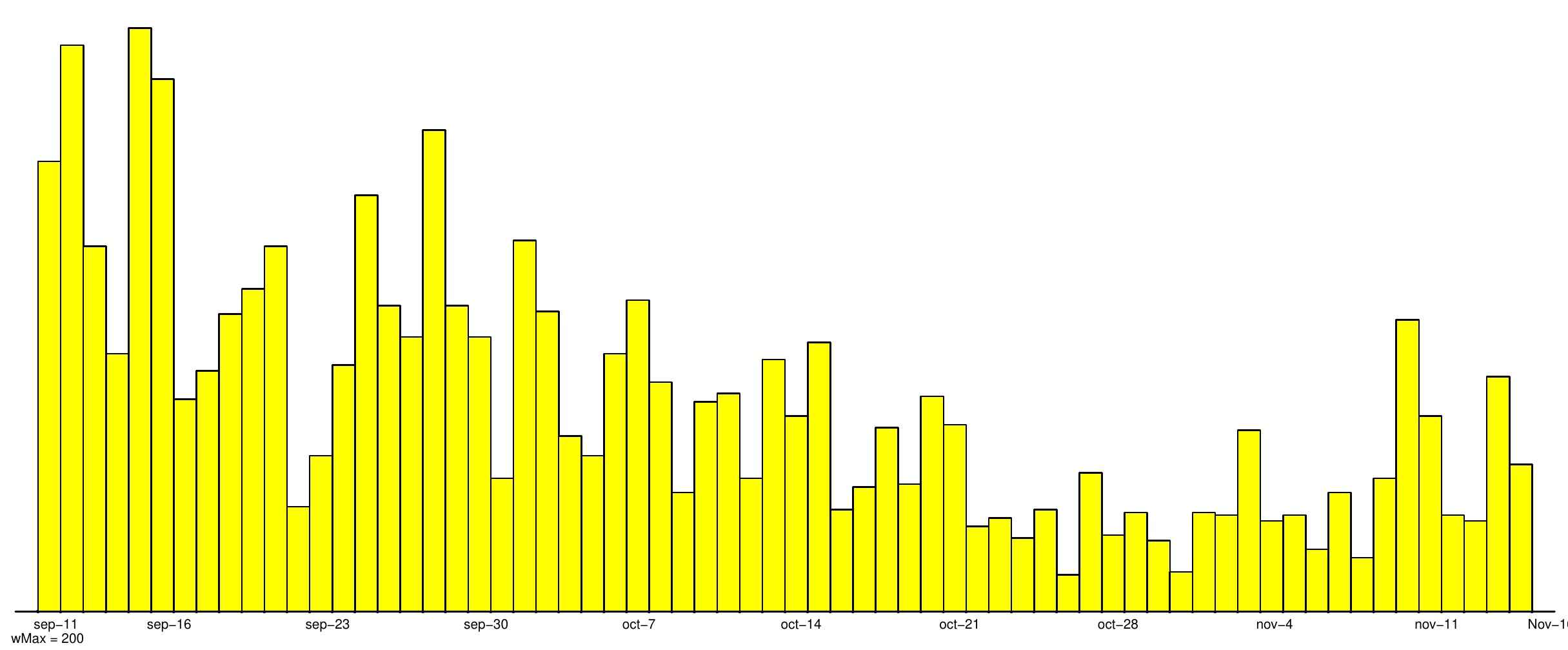}\\
  taliban :\\ 
  \includegraphics[width=70mm,viewport=80 165 780 485,clip=]{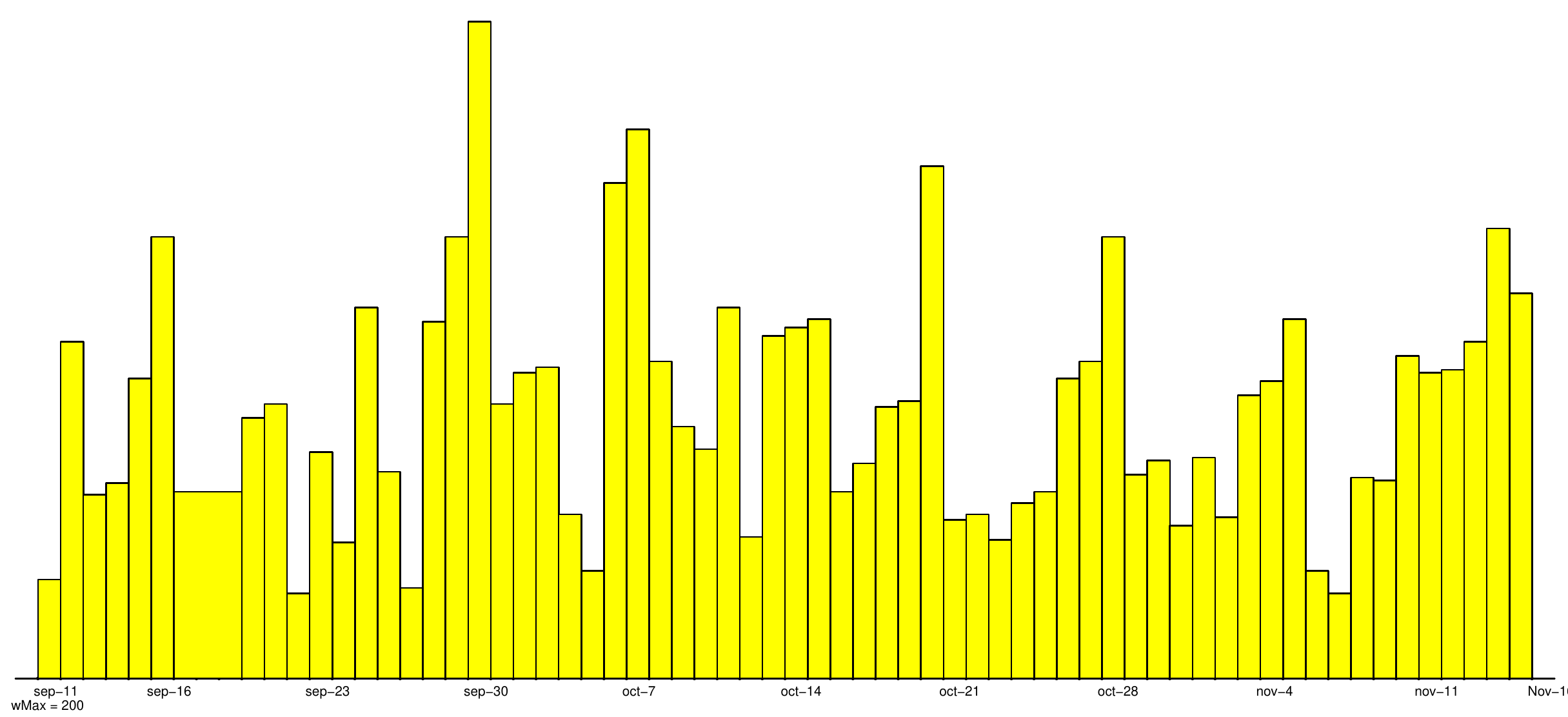}\\
  bomb :\\
  \includegraphics[width=70mm,viewport=80 165 780 320,clip=]{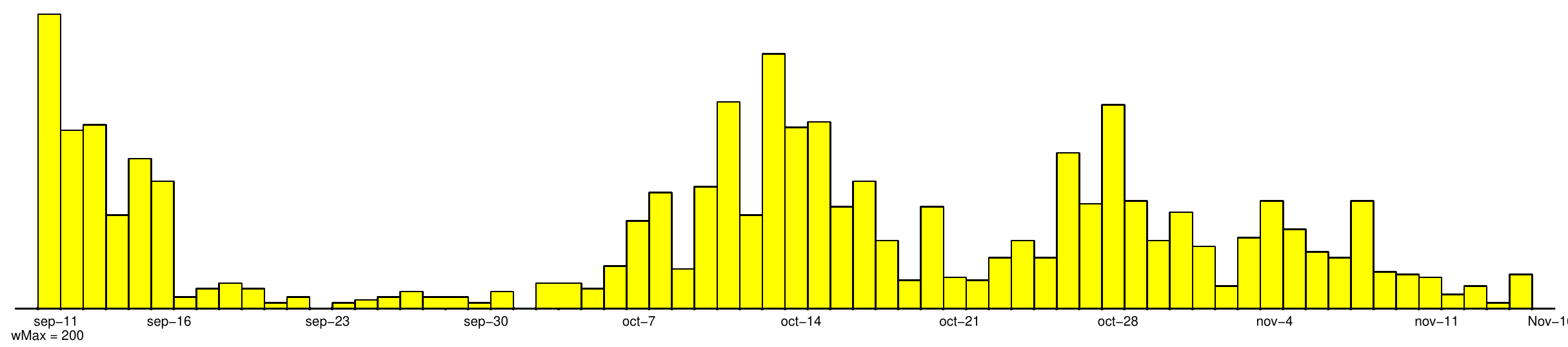}\\
  Wednesday :\\ 
  \includegraphics[width=70mm,viewport=80 165 780 540,clip=]{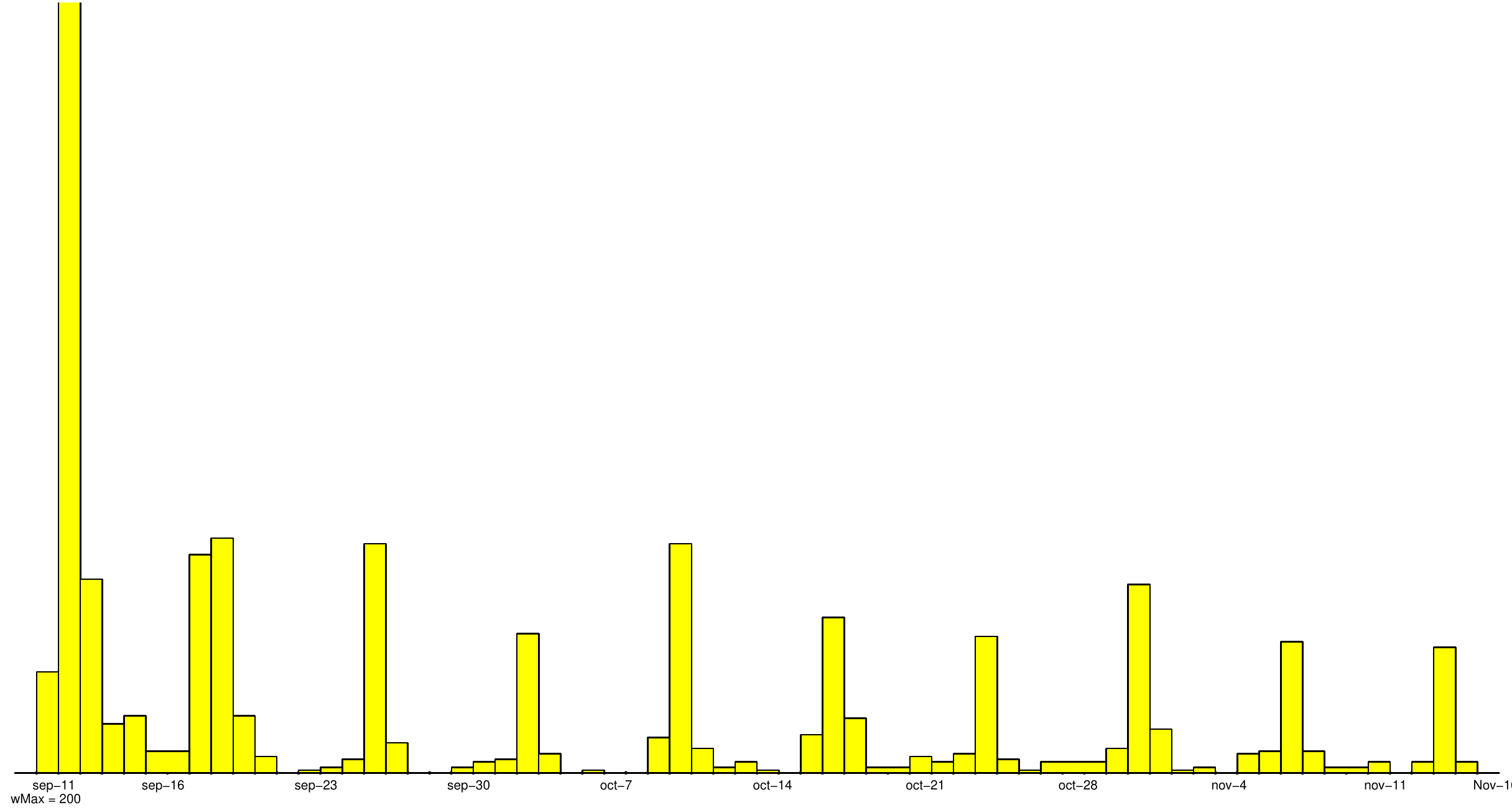}\\
   anthrax :\\
   \includegraphics[width=70mm,viewport=80 165 780 425,clip=]{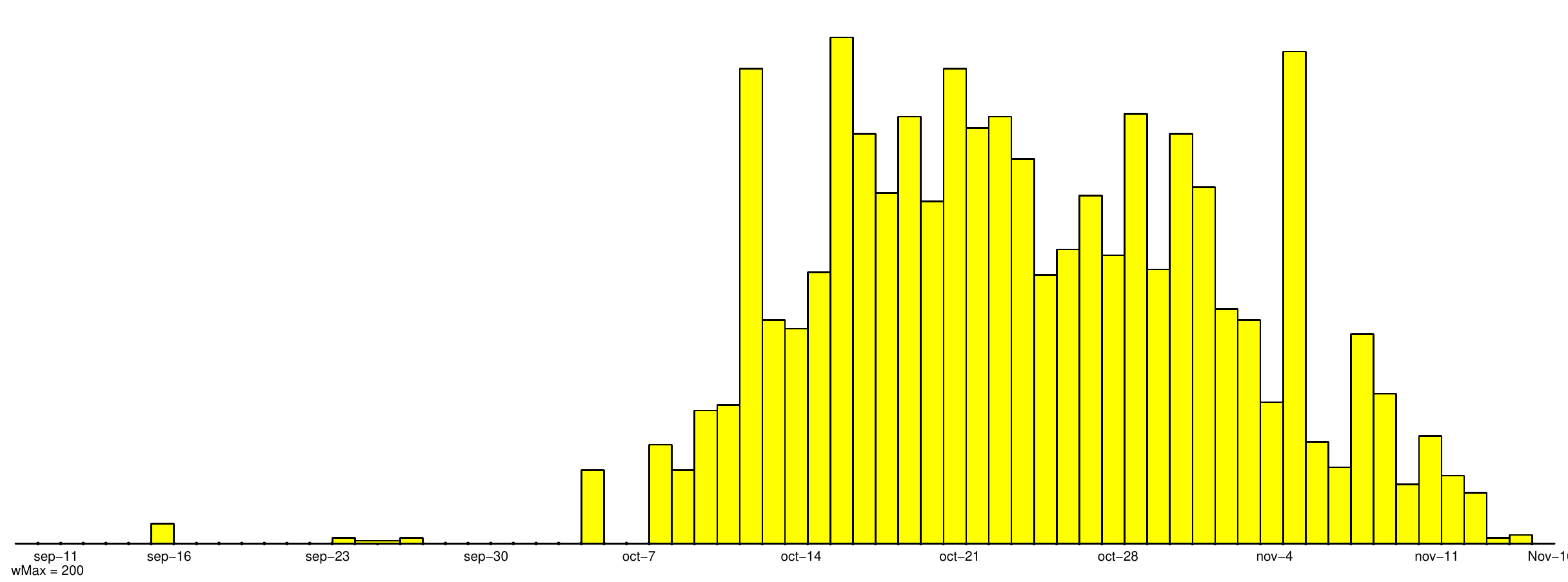}
    \end{tabular}
  \caption{Types of activity.\label{terrTy}}
 \end{center}
\end{figure}

For each of the 50 nodes we determined its temporal activity and
drew it. By visual inspection we identified 6 typical activity patterns --
types of terms (see Figure~\ref{terrTy}). For all charts in the figure
the displayed values are in the interval $[0,200]$ -- the largest
activity value for the term Wednesday is larger than 200.

\begin{table}
\caption{50 most frequent terms in the Terror news network.\label{terrF}}
\begin{center}
\begin{tabular}{r|l|r||r|l|r|}
 n &  term  & $\Sigma$freq &  n &  term  & $\Sigma$freq \\ \hline
 1 &  united\_states    &   15000	 & 	 26 &  terrorism    &    2212   \\
 2 &  attack           &   10348	 & 	 27 &  day          &    2128   \\
 3 &  taliban          &    6266	 & 	 28 &  week         &    2017   \\
 4 &  people           &    5286	 & 	 29 &  worker       &    1983   \\
 5 &  afghanistan      &    5176	 & 	 30 &  office       &    1967   \\
 6 &  bin\_laden        &    4885	 & 	 31 &  group        &    1966   \\
 7 &  new\_york         &    4832	 & 	 32 &  air          &    1962   \\
 8 &  pres\_bush        &    4506	 & 	 33 &  minister     &    1919   \\
 9 &  washington       &    4047	 & 	 34 &  time         &    1898   \\
10 &  official         &    3902	 & 	 35 &  hijack       &    1884   \\
11 &  anthrax          &    3563	 & 	 36 &  strike       &    1818   \\
12 &  military         &    3394	 & 	 37 &  afghan       &    1775   \\
13 &  plane            &    3078	 & 	 38 &  flight       &    1775   \\
14 &  world\_trade\_ctr  &    3006	 & 	 39 &  tell         &    1746   \\
15 &  security         &    2906	 & 	 40 &  terrorist    &    1745   \\
16 &  american         &    2825	 & 	 41 &  airport      &    1741   \\
17 &  country          &    2794	 & 	 42 &  pakistan     &    1714   \\
18 &  city             &    2689	 & 	 43 &  tower        &    1685   \\
19 &  war              &    2679	 & 	 44 &  bomb         &    1674   \\
20 &  tuesday          &    2635	 & 	 45 &  new          &    1650   \\
21 &  pentagon         &    2620	 & 	 46 &  buildng      &    1634   \\
22 &  force            &    2516	 & 	 47 &  wednesday    &    1593   \\
23 &  government       &    2380	 & 	 48 &  nation       &    1589   \\
24 &  leader           &    2375	 & 	 49 &  police       &    1587   \\
25 &  world            &    2213	 & 	 50 &  foreign      &    1558   \\
\hline
\end{tabular}
\end{center}
\end{table}

The \emph{primary} terms are the terms with a very high frequency of appearance in
the first week after September 11th and smaller, slowly declining
values in the following period. The representative of this group 
in Figure~\ref{terrTy} is \textbf{hijack} and other members are:
airport, american, attack, city, day, flight, nation,
New York, official, Pentagon, people, plane, police, president Bush, security, 
tower, United States, Washington,  world, World Trade center.
These are the terms describing the event.

The \emph{secondary} terms are a reaction to the event. There are
no big changes in their values. We identified three subgroups: 
a) \emph{slowly declining} represented with \textbf{bin Laden}
 (country, foreign, government, military, minister, new, 
 Pakistan, tell, terrorism, terrorist, time, war, week); 
b) \emph{stationary} represented with 
\textbf{taliban} (afghan, Afghanistan, force, group, leader); and
c) \emph{occasional} with several peaks, represented with 
\textbf{bomb} (air, building, office, strike, worker).

There are three special patterns -- two \emph{periodic}
\textbf{Wednesday} and Tuesday; and one \emph{episodic} \textbf{anthrax}.
 
To consider in a measure of importance of the node $u \in \vertices{V}$ 
also the node's position in the network  we constructed the
attraction coefficient $\mbox{att}(u)$.

Let $\mathbf{A} = [ a_{uv}]$  be a network matrix of temporal quantities with
positive real values.  We define the \keyw{node activity}
$\mbox{act}(u)$ as (see Section~\ref{activ})
\[ \mbox{act}(u) = \mbox{act}(\{u\},\vertices{V}\setminus \{u\}) = \sum_{v \in \vertices{V}\setminus \{u\}} a_{uv} .\]
Then the \keyw{attraction} of the node $u$ is defined as
\[ \mbox{att}(u) = \frac{1}{\Delta} \sum_{v \in \vertices{V}\setminus \{u\}}
            \frac{a_{vu}}{\mbox{act}(v)}  . \]
Note that the fraction $\frac{a_{vu}}{\mbox{act}(v)}$ is measuring the
proportion of the activity of the node $v$ that is shared with the node $u$.

From $0 \leq \frac{a_{vu}}{\mbox{act}(v)} \leq 1$ and $\deg(v)=0 \Rightarrow a_{vu}=0$
it follows that
\[ \sum_{v \in \vertices{V}\setminus \{u\}}
            \frac{a_{vu}}{\mbox{act}(v)} \leq \deg(u) \leq \Delta  \]
where $\Delta$ denotes the maximum degree.            
Therefore we have $0 \leq \mbox{att}(u) \leq 1$, for all $u \in \vertices{V}$.

The maximum possible attraction value 1 is attained exactly for nodes:
a) in an undirected network: that are the root of a star;
b) in a directed network: that are the only out-neighbors of their in-neighbors --
the root of a directed in-star.

\begin{figure}[!]
 \begin{center}
    \begin{tabular}{l}
   pres Bush :\\
   \includegraphics[width=70mm,viewport=80 165 780 535,clip=]{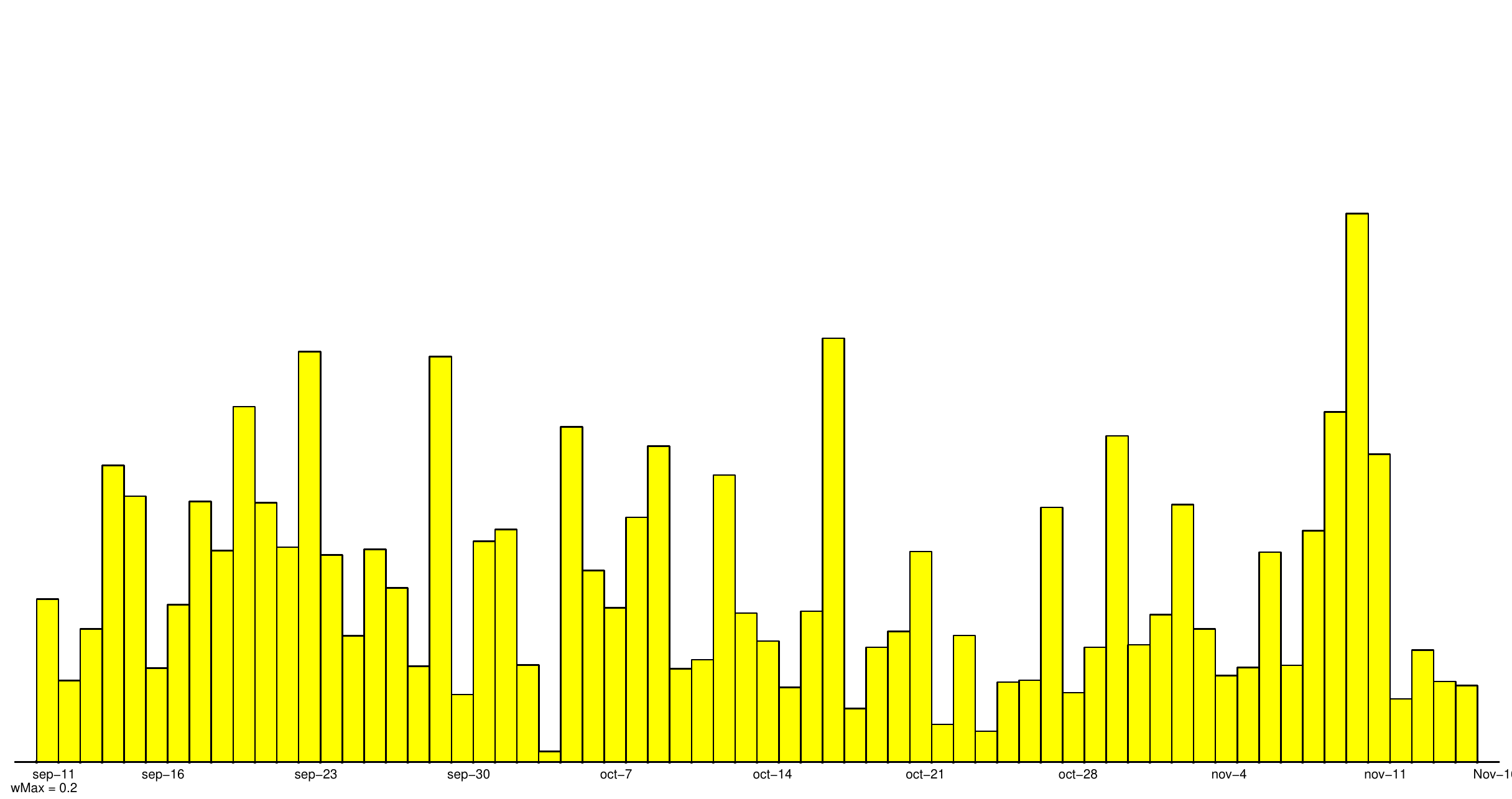}\\
   Pakistan :\\
   \includegraphics[width=70mm,viewport=80 165 780 415,clip=]{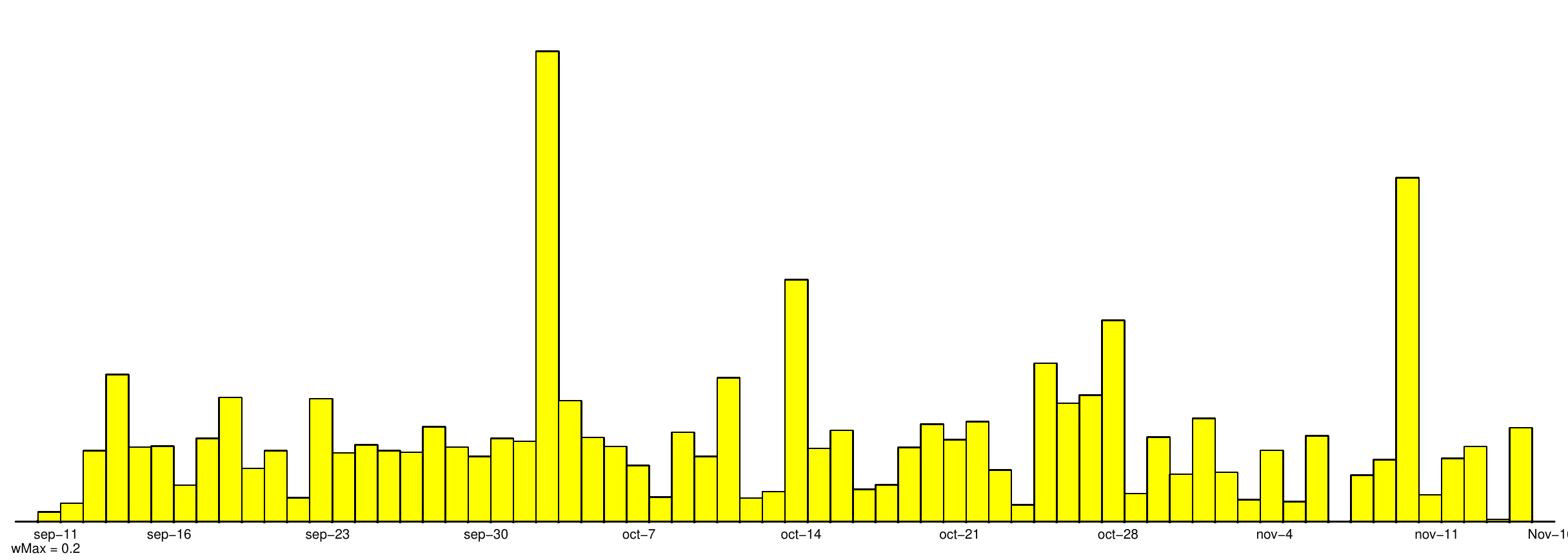}\\
   taliban :\\ 
  \includegraphics[width=70mm,viewport=80 165 780 520,clip=]{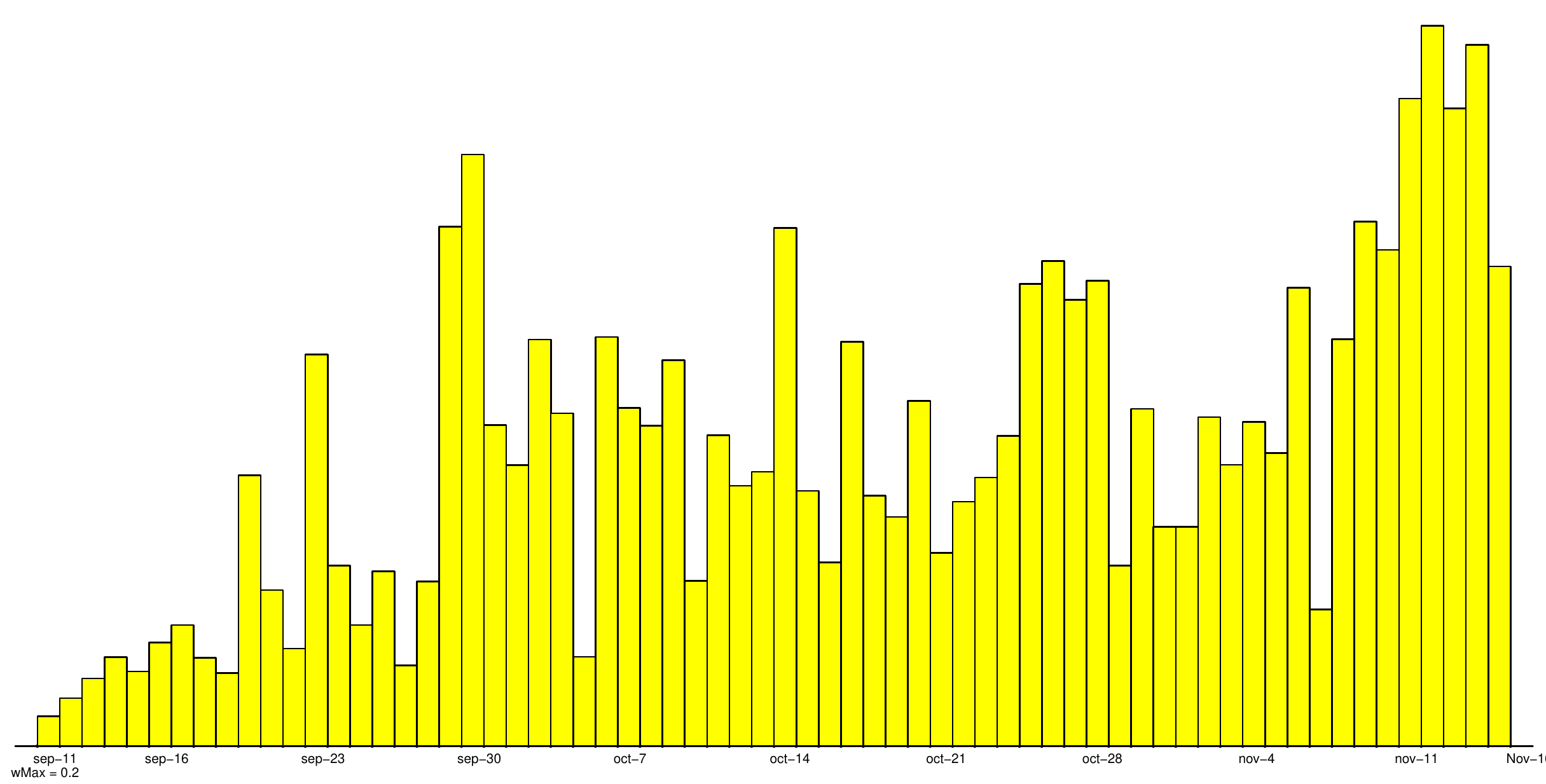}\\
   Kabul :\\
  \includegraphics[width=70mm,viewport=80 165 780 430,clip=]{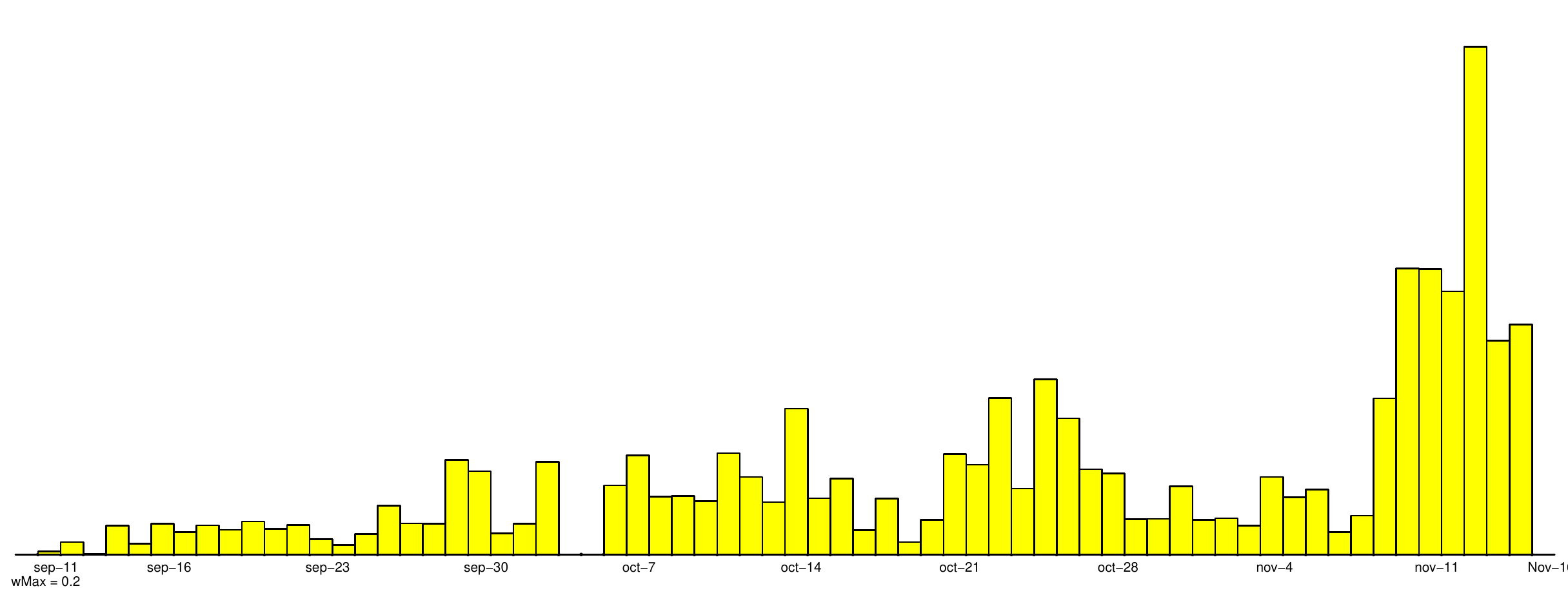}\\
   bomb :\\
  \includegraphics[width=70mm,viewport=80 165 780 360,clip=]{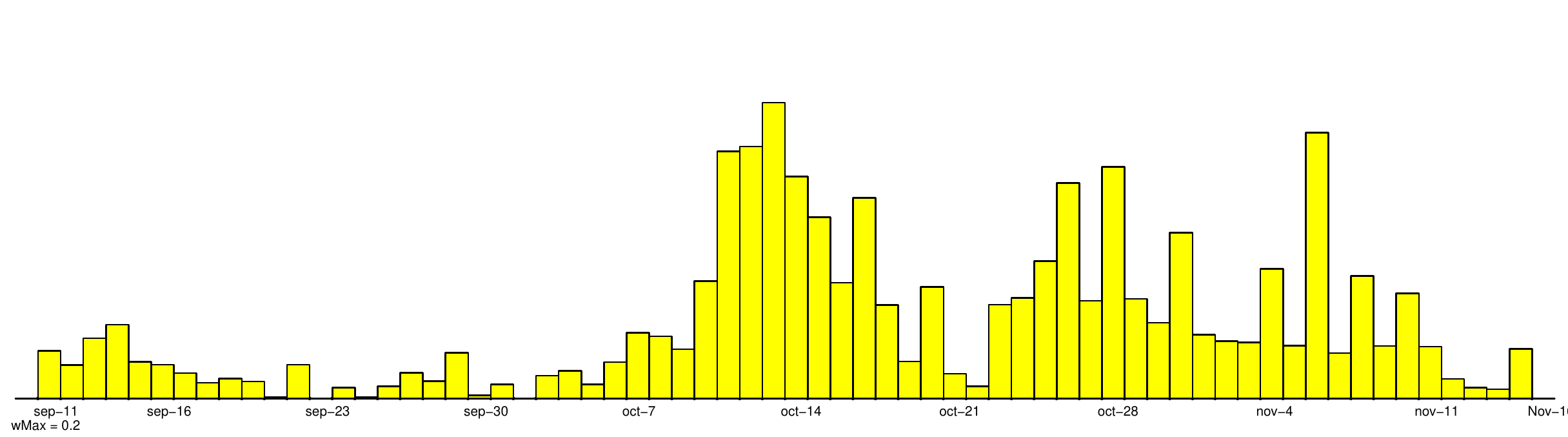}\\
  anthrax :\\
   \includegraphics[width=70mm,viewport=80 165 780 495,clip=]{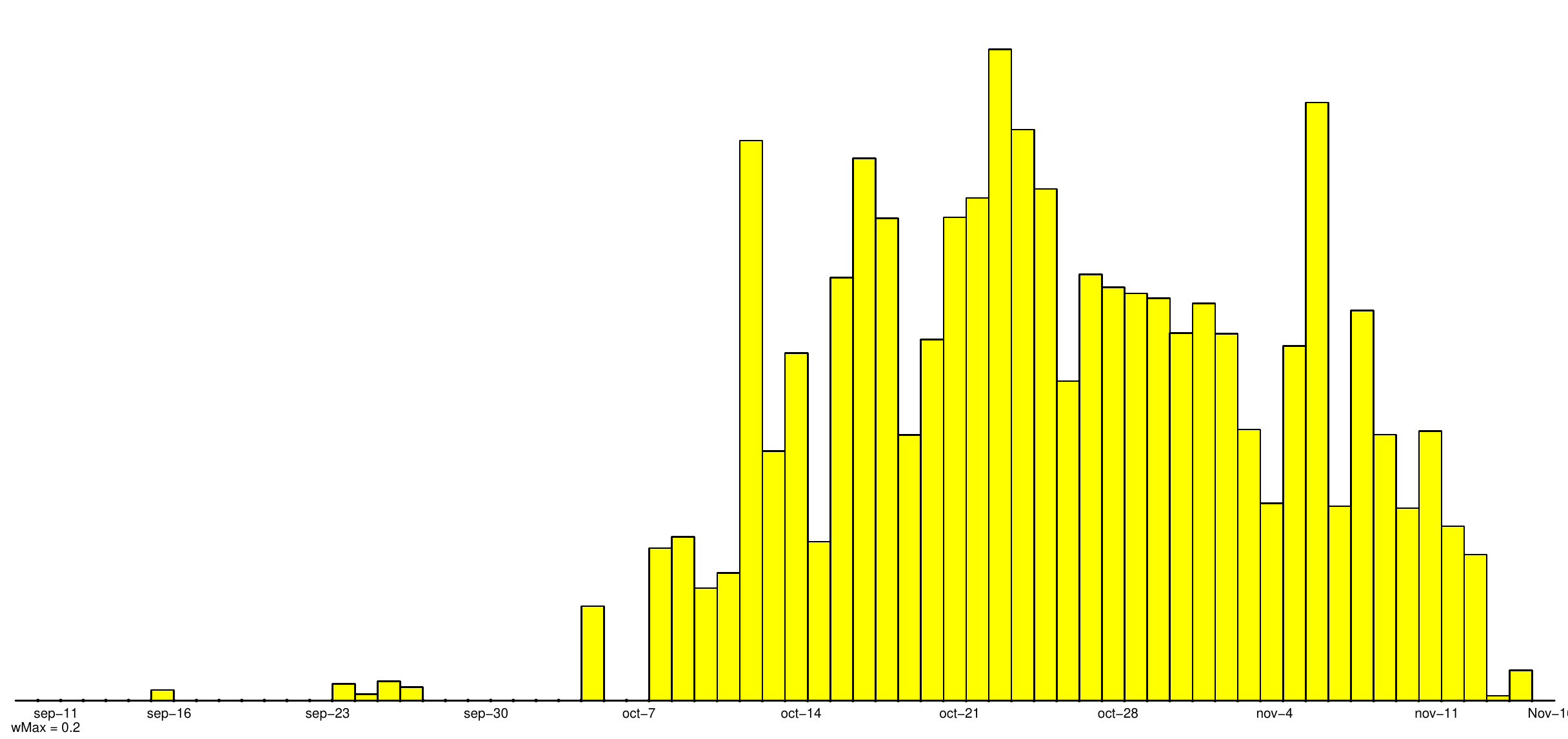}
    \end{tabular}
  \caption{Attraction patterns.\label{terrAt}}
 \end{center}
\end{figure}

We computed the temporal attraction and the corresponding aggregated attraction
values for all the nodes in our network. We selected 30 nodes with the
largest aggregated attraction values. They are listed in Table~\ref{terrA}.
Again we visually explored them. In Figure~\ref{terrAt} we present 
temporal attraction coefficients for the 6 selected terms. For all charts in the figure
the displayed attraction values are in the interval $[0,0.2]$.

Comparing on the common terms (taliban, bomb, anthrax) the activity
charts in Figure~\ref{terrTy} with the corresponding attraction charts in
Figure~\ref{terrAt} we see that they are ``correlated'' (obviously 
$\mbox{act}(a;t) = 0$ implies $\mbox{att}(a;t) = 0$), but different
in details.

For example, the terms taliban and bomb have small attraction values at
the beginning of the time window -- the terms were disguised by the
primary terms. On the other hand, the terms taliban and Kabul get
increased attraction towards the end of the time window.
 
\begin{table}
\caption{30 most attractive terms in the Terror news network.\label{terrA}}
\begin{center}
\begin{tabular}{r|l|r||r|l|r|}
 n &  term  & $\Sigma$att &  n &  term  & $\Sigma$att \\ \hline
 1 &  united\_states   &     12.216	 &    16 &  war           &        2.758   \\ 
 2 &  taliban         &      7.096	 &    17 &  force         &        2.596   \\ 
 3 &  attack          &      7.070	 &    18 &  new\_york      &        2.590   \\ 
 4 &  afghanistan     &      5.142	 &    19 &  government    &        2.496   \\ 
 5 &  people          &      5.023	 &    20 &  day           &        2.338   \\ 
 6 &  bin\_laden       &      4.660	 &    21 &  leader        &        2.305   \\ 
 7 &  anthrax         &      4.601	 &    22 &  terrorism     &        2.202   \\ 
 8 &  pres\_bush       &      4.374	 &    23 &  time          &        2.182   \\ 
 9 &  country         &      3.317	 &    24 &  group         &        2.072   \\ 
10 &  washington      &      3.067	 &    25 &  afghan        &        2.040   \\ 
11 &  security        &      2.939	 &    26 &  world         &        1.995   \\ 
12 &  american        &      2.922	 &    27 &  week          &        1.961   \\ 
13 &  official        &      2.831	 &    28 &  pakistan      &        1.943   \\ 
14 &  city            &      2.798	 &    29 &  letter        &        1.866   \\ 
15 &  military        &      2.793	 &    30 &  new           &        1.851   \\ 
\hline
\end{tabular}
\end{center}
\end{table}

\section{Conclusions}

In the paper we proposed an algebraic approach to the ``deterministic''
analysis of temporal networks based on temporal quantities and presented algorithms
for the temporal variants of basic network analysis measures and concepts.
We expect that  the support for many temporal variants of 
other network analysis notions can be developed in similar ways.
Our results on temporal variants of eigen values/vectors based indices
(Katz, Bonacich, hubs and authorities, page rank) are presented in a
separate paper \citep{tcent}.

The proposed approach is an alternative to the traditional
cross sectional approach based on time slices. Its main advantages are:
\begin{itemize}
\item the data and the results are expressed using temporal quantities that
are natural descriptions of properties changing through time;
\item the user does not need to be careful about the intervals on which the
time slices are determined -- exactly the right intervals are selected by the
merging (sub)operations. This also improves, on average, the efficiency of
the proposed algorithms.
\end{itemize}

All the described algorithms (and some others) are implemented in a Python
library TQ (Temporal Quantities) available at the page
\cite{TQ}.
We started to develop a program Ianus
that will provide a user-friendly (Pajek like) access to the capabilities
of the TQ library.

The main goal of the paper was to show: it can be done. Therefore we
based the current version of the library TQ on a matrix representation of
temporal networks as it is presented in the paper. For this representation
most of the network algorithms have the time complexity of $O(n^3 \cdot L)$ and
the space complexity of $O(n^2 \cdot L)$. This implies that their application is
limited to networks of moderate size (up to some thousands of nodes).
Large networks are usually sparse. On this assumption more efficient
algorithms can be developed based on a graph (sparse matrix) representation --
one of the directions of our future research.

In a description of a temporal network $\network{N}$ we can consider also a transition
time or latency $\tau \in \functions{W}$: $\tau(l,t)$ is equal to the time needed to traverse the
link $l$ starting at the instant $t$. Problems considering latency are typical for 
operations research but could be important, when such data are available, also in
social network analysis \citep{moody,Xuan,Rout,TVGsur,alge}. The analysis of temporal networks
considering also the latency seems a much harder task -- for example, in such temporal 
networks the strongly connected components problem is NP-complete \citep{SCC}.

The results obtained from temporal procedures are relatively large. To
identify interesting elements we used in the paper the aggregated values
and the visualization of selected elements. Additional tools for browsing
and presenting the results should be developed.
 
\begin{acknowledgments}
	
The work was supported in part by the ARRS, Slovenia, grant J5-5537, as well as
by a grant within the EUROCORES Programme EUROGIGA (project GReGAS) of the European Science Foundation.

\end{acknowledgments}

\FloatBarrier

\end{document}